\documentclass[acmsmall,screen,nonacm]{acmart}
\usepackage{libertine}
\usepackage{booktabs}
\usepackage{tabularx}
\usepackage{multirow}
\usepackage{listings}
\usepackage{xtab}
\usepackage{caption}
\usepackage{algorithm}
\usepackage[noend]{algpseudocode}
\usepackage{tikz}
\usepackage{balance}
\PassOptionsToPackage{hyphens}{url}\usepackage{hyperref}

\usepackage{wrapfig}
\usepackage{multicol}
\usepackage[framemethod=tikz]{mdframed}
\usepackage{needspace}

\usepackage{xcolor}
\definecolor{codegreen}{rgb}{0.0,0.5,0.0}
\definecolor{codegray}{rgb}{0.4,0.4,0.4}
\definecolor{codepurple}{rgb}{0.5,0.0,0.5}
\definecolor{codeblue}{rgb}{0.0,0.3,0.6}
\definecolor{codeorange}{rgb}{0.8,0.4,0.0}
\definecolor{backcolour}{rgb}{0.97,0.97,0.97}
\definecolor{framecolour}{rgb}{0.85,0.85,0.85}

\definecolor{algobg}{rgb}{0.95,0.97,1.0}
\definecolor{algoborder}{rgb}{0.3,0.5,0.8}
\definecolor{algokeyword}{rgb}{0.0,0.3,0.7}
\definecolor{algofunction}{rgb}{0.5,0.0,0.5}
\definecolor{algocomment}{rgb}{0.0,0.5,0.2}

\lstdefinestyle{mystyle}{
    backgroundcolor=\color{backcolour},   
    commentstyle=\color{codegreen}\itshape,
    keywordstyle=\color{codeblue}\bfseries,
    numberstyle=\tiny\color{codegray},
    stringstyle=\color{codeorange},
    basicstyle=\ttfamily\small,
    breakatwhitespace=false,         
    breaklines=true,                 
    captionpos=b,                    
    keepspaces=true,                 
    numbers=none,                    
    numbersep=8pt,                  
    showspaces=false,                
    showstringspaces=false,
    showtabs=false,                  
    tabsize=2,
    frame=none,
    xleftmargin=3pt,
    xrightmargin=3pt,
    aboveskip=4pt,
    belowskip=2pt,
    columns=flexible
}
\mdfdefinestyle{listingstyle}{
    linecolor=framecolour,
    linewidth=1pt,
    backgroundcolor=backcolour,
    innertopmargin=3pt,
    innerbottommargin=3pt,
    innerleftmargin=3pt,
    innerrightmargin=3pt,
    roundcorner=2pt,
    skipabove=\topsep,
    skipbelow=\topsep,
    nobreak=false
}

\makeatletter
\renewcommand{\ALG@name}{Algorithm}

\mdfdefinestyle{algorithmstyle}{
    linecolor=algoborder,
    linewidth=1.2pt,
    backgroundcolor=algobg,
    innertopmargin=2pt,
    innerbottommargin=5pt,
    innerleftmargin=5pt,
    innerrightmargin=5pt,
    roundcorner=2pt,
    skipabove=\topsep,
    skipbelow=\topsep,
    nobreak=false
}

\AtBeginDocument{%
  }

\setcopyright{None}

\begin{document}

\title{Modeling Layout Abstractions Using Integer Set Relations}

\author{Somashekaracharya Gunasagara Bhaskaracharya}
\email{sbhaskaracha@nvidia.com}
\affiliation{%
  \institution{NVIDIA}
  \country{India}
}

\author{Aravind Acharya}
\email{aravinda@nvidia.com}
\affiliation{%
  \institution{NVIDIA}
  \country{India}
}

\author{Bastian Hagedorn}
\email{bhagedorn@nvidia.com}
\affiliation{%
  \institution{NVIDIA}
  \country{Germany}
}

\author{Vinod Grover}
\email{vgrover@nvidia.com}
\affiliation{%
  \institution{NVIDIA}
  \country{USA}
}

%
\begin{abstract}
	Modern deep learning compilers rely on layout abstractions to manage the complex
mapping between logical tensor structures and physical memory arrangements. CuTe
layouts and Triton linear layouts are widely adopted industry standards.
However, these layout systems operate independently with distinct mathematical
underpinnings, preventing unified formal analysis and cross-system reasoning. We
bridge this gap by introducing a novel approach that leverages the Integer Set
Library (ISL) to create a unified mathematical representation for both layout
systems through integer set relations, thereby enabling rigorous formal
analysis, correctness verification, and the foundation for future cross-system
optimization strategies.

Our approach models CuTe layouts through integer set relations that encode the
transformation from multi-dimensional coordinates to linear indices using
stride-based calculations, including sophisticated swizzle operations that
perform bit-level manipulations for enhanced memory access patterns. For Triton
linear layouts, we construct integer set relations that model the binary vector
space transformations where arithmetic operations follow finite field $F_2$
rules. We implement a complete suite of layout manipulation algorithms for
composition, inversion, complement using built-in operations in ISL to
ensure mathematical correctness and preserve layout semantics. Experimental
evaluation shows that the system handles the full spectrum of layout complexity,
from elementary identity transformations to sophisticated multi-dimensional
tensor arrangements with complex stride configurations and swizzle patterns,
validating the mathematical modeling approach across different layout paradigms.

This work provides the theoretical groundwork for consolidating layout
abstraction methodologies in deep learning compilation. The resulting framework
enables formal verification of layout transformations, comparative analysis
across different layout systems, facilitates integration with established
polyhedral compilation infrastructures, and creates the foundation for
developing sophisticated automatic optimization techniques critical for
advancing deep learning compilation technology.


\end{abstract}

%
%

\maketitle

\section{Introduction}
The rapid evolution of deep learning has fundamentally transformed the landscape
of computational workloads, with modern neural networks requiring unprecedented
computational resources and sophisticated optimization strategies.
Modern GPUs feature specialized matrix multiplication instructions that require specific data layouts for optimal performance.
However, these compute-optimized layouts are often inefficient for memory operations, necessitating data rearrangement at multiple levels.
Layout abstractions address this challenge by defining tensor mappings between logical coordinate spaces and physical memory locations across the GPU's compute and memory hierarchy.
These abstractions are fundamental components in deep learning 
compilers~\cite{tillet2019triton, cutedsl, graphene_asplos} and fast kernel
libraries~\cite{cutlass} that express mappings between hardware resources and
logical tensor indices, playing a crucial role in optimizing memory access
patterns and hardware utilization in deep learning workloads.

Deep learning frameworks integrate layout abstractions to enhance performance of
GPU-accelerated operations, particularly in kernels involving matrix
multiplication, convolution, and attention mechanisms. Two prominent layout
abstraction systems have emerged as industry standards: CuTe layouts~\cite{cute}
and Triton linear layouts~\cite{zhou2025linearlayoutsrobustcode, triton_linear_layout}. These systems represent
fundamentally different approaches to the same underlying problem of 
tensor layout management and transformation.

CuTe layouts specify mappings from $n$-D coordinate spaces to
1-D index spaces using shape and stride tuples. For example, a CuTe
layout with shape $(3,4)$ and strides $(4,1)$ maps logical coordinates $(i,j)$
to linear index $4i + j$. These layouts are used to facilitate efficient memory
access and data organization in deep learning operations, providing a flexible
system for expressing tensor layouts and transformations. CuTe layouts have
become integral to NVIDIA's CUTLASS library~\cite{cutlass}, enabling
high-performance linear algebra operations through sophisticated layout
manipulation. Additionally, CuTe supports swizzle operations that perform
bit-level manipulations on offsets and pointers to optimize memory access patterns for
specific hardware configurations, further enhancing performance through
sophisticated address space transformations.  In contrast, linear layouts, used
in OpenAI Triton~\cite{triton}, offer a different approach based on binary
vector spaces and mappings between them. Linear layouts map $m$-D
coordinate spaces to $n$-D index spaces, relying on the concept of
binary vector spaces where addition corresponds to an XOR operation and
multiplication to an AND operation. Both these approaches provide a mathematical
foundation for understanding how hardware resources and tensor positions are
mapped to each other. In general, the input and output spaces involved in these
mappings can correspond to logical tensor positions or hardware resource indices
e.g., thread indices.


Despite their widespread adoption and proven effectiveness, these layout
abstraction systems exist in isolation, each with their own mathematical
foundations and implementation approaches. This fragmentation creates several
significant challenges for the deep learning compiler ecosystem. Firstly, the
lack of a unified mathematical framework makes it difficult to perform formal
analysis and automatic optimization across different systems.  Compiler
developers must implement separate analysis and optimization passes for each
layout abstraction, leading to code duplication and maintenance overhead.
Secondly, the absence of a common mathematical language hinders the development
of cross-system optimization techniques that could leverage the strengths of
multiple layout abstraction approaches. Thirdly, the lack of formal verification
capabilities makes it challenging to ensure correctness of layout
transformations, particularly when complex optimizations are applied.


The Integer Set Library (ISL)~\cite{isl} provides a powerful framework for
representing and manipulating integer sets and relations, making it well-suited
for modeling layout mappings. ISL has been extensively used in polyhedral
compilation.  The polyhedral model has proven to be a powerful tool for
analyzing and optimizing affine loop nests, capturing iteration spaces, array
spaces, array accesses, and schedules as integer polyhedra and relations between
them. Advanced polyhedral compilation techniques can then be employed to analyze
and transform these loop nests as required. Polyhedral compilation techniques
have demonstrated their effectiveness in optimizing loop nests both in the
context of domain-specific~\cite{baghdadi2019cgo, mullapudi2015asplos} and
general-purpose~\cite{uday08pldi, polly, verdoolaege2013taco, rstream}
compilation. In the context of deep learning, polyhedral techniques have been
successfully applied in frameworks like Tensor
Comprehensions~\cite{vasilache2018tcCoRR}, MLIR~\cite{mlircgo2021}, and
Diesel~\cite{elango2018mapl, bhaskaracharya2020arxiv}.  These frameworks
demonstrate the effectiveness of polyhedral analyses for optimizing neural
network computations.

The motivation for this work stems from the need to address a fundamental
research question: \textit{is there a unified mathematical formulation that
consolidates the various layout abstractions currently in use?} We propose
integer set relations as this unifying framework. To the best of our
knowledge, we are the first to establish this connection. This work is
fundamentally theoretical and foundational in nature. Our goal is not to
provide performance optimizations, but rather to lay the mathematical groundwork
that enables formal reasoning about diverse layout abstractions. 
Integer set relations are more expressive than the mappings typically used in
existing layout systems. By expressing layouts as integer set relations, we can
leverage the capabilities of ISL for formal analysis, optimization, and
verification of layout transformations.
Since ISL is widely used in compiler toolchains~\cite{polly,
uday08pldi, pet} that employ polyhedral modeling of static control parts
(SCoPs)~\cite{clan_bastoul} to analyze and optimize them, as well as in deep
learning compilers~\cite{vasilache2018tcCoRR, mlircgo2021, elango2018mapl} that
rely on polyhedral analyses and optimizations, establishing a connection between
various layout abstractions and integer set relations is crucial for
enabling advanced compiler optimizations and formal verification. This paper
makes the following key contributions to the field of deep learning compilation:




\begin{itemize}

\item Unified Mathematical Framework for Layout Abstractions: We present the
first framework to model both CuTe layouts (including swizzle operations with
bit-level manipulations like XOR and bit-shifting) and Triton linear layouts
using a unified mathematical representation that is based on integer set
relations in ISL, thereby enabling formal analysis and optimization across
previously isolated layout systems.

\item Layout Operation Algorithms: We show that the modeling 
is versatile enough to develop algorithms for transforming CuTe
layouts using built-in ISL operations while preserving their semantic properties
and ensuring correctness. This includes operations such as composition, inversion
and complement which enable more complex layout operations.

\item Practical Implementation: We demonstrate the effectiveness of our
framework through the \texttt{isl-layout} tool, which successfully translates
various layout specifications into integer set relations and handles layouts of
varying complexity, providing concrete examples of the mathematical modeling
approach in practice.

\end{itemize}

The remainder of this paper is organized as follows.
Section~\ref{sec:background} provides comprehensive background on CuTe and
linear layout abstractions as well as on ISL and the polyhedral model.
Section~\ref{sec:cute} describes in detail how CuTe layouts can be represented
as integer set relations, including algorithms for performing layout operations
using ISL. Section~\ref{sec:cute-swizzles} discusses the modeling of CuTe
swizzles using ISL.  Section~\ref{sec:ll} presents our framework for
representing linear layouts in Triton as integer set relations between binary
coordinate and index spaces. Section~\ref{sec:implementation} demonstrates the
\begin{figure}[t]
    \centering
    \includegraphics[width=0.4\textwidth]{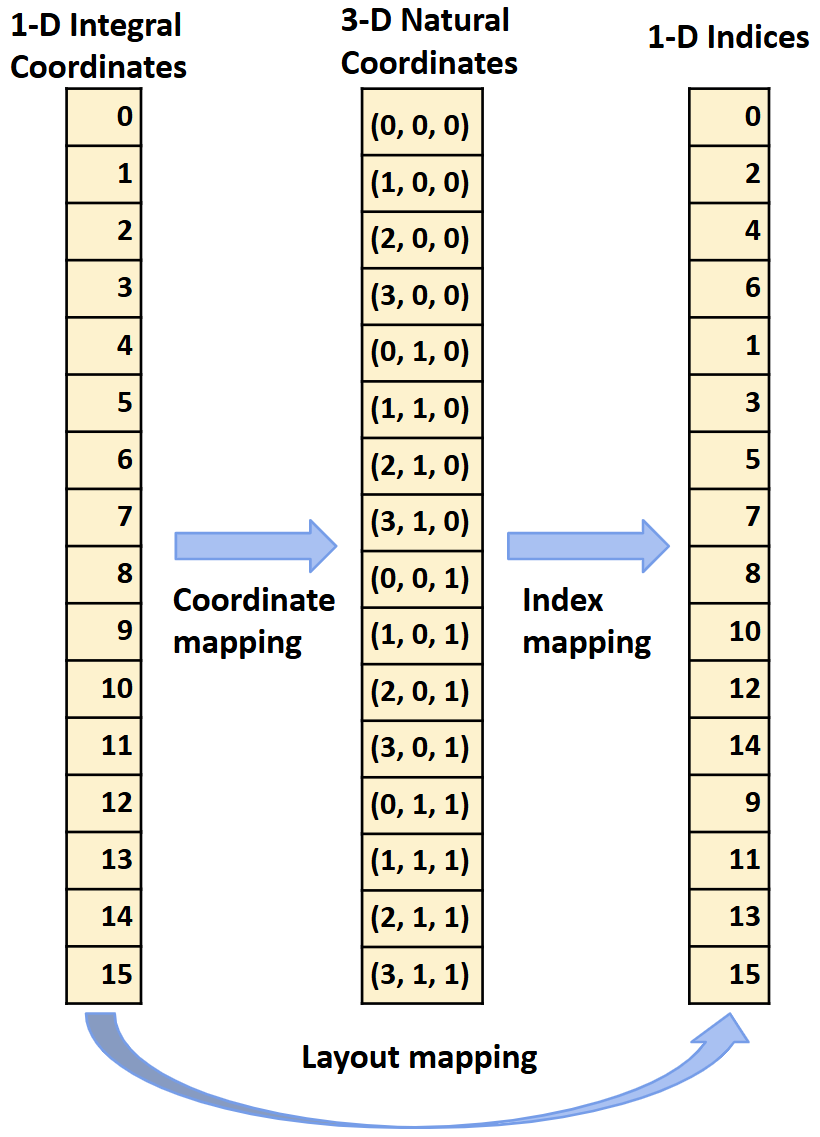}
    \caption{Mappings defined by the CuTe layout $H = (4, 2, 2) : (2, 1, 8)$}
    \label{fig:cute-layout-mapping}
\end{figure}
practical implementation through various example layouts and layout operations.
Section~\ref{sec:related} discusses related work in polyhedral compilation, deep
learning frameworks, and GPU optimization.
Finally,~Section~\ref{sec:conclusions} concludes with a summary of our
contributions and outlines future directions for research.

\section{Background}
\label{sec:background}
In this section we provide the necessary background on CuTe and linear
layouts, ISL and the polyhedral model.

\subsection{CuTe Layout}
\label{sec:cute-layout}
A CuTe layout $s : d$ specifies mappings from $n$-D coordinate spaces to a 1-D
index space, where $s$ is a tuple that represents the shape of the $n$-D
coordinate space and $d$ is a tuple of equal rank specifying the strides of the
layout. For example, consider the layout $H = (4, 2, 2) : (2, 1, 8)$. The shape
of the layout $(4, 2, 2)$ implies that it is of size 16 i.e., there are 16
points in the coordinate space. These points can be associated with a 1-D
coordinate space consisting of integer points from 0 to 15. This is referred to
as the \textit{integral coordinate} space. Furthermore, an equivalent 3-D
coordinate space, called the \textit{natural coordinate} space, is defined by
the shape $(4, 2, 2)$. 

Following a colexicographic ordering of the points in this natural coordinate
space, we can define a mapping from the integral coordinate space to the natural
coordinate space as shown in Figure~\ref{fig:cute-layout-mapping}. We refer to
this as the \textit{coordinate mapping}.  The dot product of a natural
coordinate and the strides gives the 1-D index corresponding to it. This dot
product essentially defines the \textit{index mapping}, which maps the $n$-D
coordinate space to a 1-D index space. Finally, a composition of the coordinate
mapping and the index mapping gives the overall \textit{layout mapping} from the
1-D integral coordinate space to the 1-D index space.

\subsubsection{Hierarchical Layouts}
\label{sec:hierarchical-layouts}

CuTe layouts support hierarchical nesting of shapes and strides, allowing for
more flexible layout specifications. For example, $H = (4, (2, 2)) : (2, (1,
8))$ is a hierarchical layout as the shape and stride tuples are nested.
Hierarchical layouts provide flexibility in using different natural coordinate
spaces for the same layout specification while maintaining the same overall
layout mapping. For example, the hierarchical layout $H$ supports not just the
1-D integral coordinate space but also the 3-D natural coordinate space of shape
$(4, 2, 2)$, as well as the 2-D natural coordinate space of shape $(4, 4)$.
The shape of the 1-D coordinate space, which is $16$, is said to be
\textit{compatible} with the shape of the 3-D natural coordinate space $(4, 2,
2)$ and the 2-D natural coordinate space $(4, 4)$. In general, a shape $s_1$ is
said to be compatible with a shape $s_2$ if their sizes are equal and all
co-ordinates of $s_1$ are valid co-ordinates of $s_2$ through such hierarchical
nesting.

\subsubsection{Layout Operations}
\label{sec:layout-operations}

CuTe supports various operations on layouts that result in transformed layouts.
The three fundamental operations are composition, inverse, and complement, which
form the basis for more complex operations like layout divisions and products.

\textit{Composition:} ($H = G \circ F$) defines a composition of layouts $F$ and
$G$ which results in a new layout $H$ that maps the coordinate space of $F$ to
the index space of $G$.

\textit{Inverse:} This operation computes the full inverse of a layout. For a layout
$H = s : d$, the inverse layout $H^{-1}$ has shape and strides that produce the
inverse mapping, effectively reversing the transformation from coordinates to
indices specified by the layout $H$.

\textit{Complement:} Given a layout $H$ and a positive integer value $D$, the
layout complement $H^\prime$ is another layout such that, if $HH^\prime$ is a
layout with the shape and strides of $H$ and $H^\prime$ concatenated together,
the layout mapping specified by $H^\prime$ maps to as many integers as possible
in the interval $[0, D)$.  For example, consider $H = (4, 2) : (1, 16)$ and $D =
32$.  The complement $H^\prime$ is $(4) : (4)$. The concatenated layout
$HH^\prime = (4, 2, 4) : (1, 16, 4)$ results in a layout mapping that maps to
all the integers in the interval $[0, 32)$. 

\subsection{CuTe Swizzle}
\label{sec:cute-swizzle}

CuTe supports another class of layouts called swizzled layouts~\cite{cute-swizzles} that perform
bit-level manipulations to optimize memory access patterns. A CuTe swizzle
$H_{b, m, s}$ is a bijective mapping that maps integers in the interval $[0,
2^{b + m + |s|})$ to integers in the same interval, where $b$, $m$ and $s$ are
all integers with $b$ and $m$ being non-negative. The parameters control the
swizzle behavior: $b$ determines the number of bits to be swizzled, $m$
specifies the number of least significant bits that remain unchanged, and $s$
defines the shift width for the swizzle operation. For an integer $c$ in this
interval, the exact definition of the swizzle is based on bit manipulation and
is given by the equation:

\begin{equation}
    H_{b, m, s}(c) = c \oplus ((c\ \&\ y) \gg s).
\end{equation}

The bit-wise XOR operation is represented by $\oplus$ while $\&$ represents the
bit-wise AND operation. The bit-shift operation is represented by the $\gg$
operator (right shift for positive $s$, left shift for negative $s$). The AND
operation essentially represents a masking operation using the mask value $y$
which is equal to $((2^{b} - 1) \ll (m + \max(s, 0)))$.

\subsection{Linear Layout}
\label{sec:linear-layout}

Linear layouts in Triton specify mappings from $m$-D coordinate spaces to $n$-D
index spaces. The key idea behind linear layouts is that the complete linear
layout can be specified using only a few basis vectors rather than enumerating
all possible mappings. For example, for a linear layout $H$ involving 2-D
coordinates $(t, w)$ and 2-D indices $(x, y)$, it suffices to define the values
of $L(t,0)$ and $L(0,w)$ where $t$ and $w$ are powers of 2. Suppose these are
$t_1, t_2$ and $w_1, w_2$ respectively. The entire layout can then be computed
using the linearity rule:
\begin{equation}
    L(t_1 \oplus t_2, w_1 \oplus w_2) = L(t_1, w_1) \oplus L(t_2, w_2),
\end{equation}
where $\oplus$ represents the bitwise XOR operation.

\subsubsection{Binary Vector Space}
\label{sec:binary-vector-spaces}
Linear layouts can be viewed as linear transformations between binary vector
spaces over the finite field $F_2$. Given an $m$-D coordinate space with
dimension sizes $s_0, s_1, \ldots, s_{m-1}$ and an $n$-D index space with
dimension sizes $s'_0, s'_1, \ldots, s'_{n-1}$, these spaces can be split into
binary vector spaces i.e., the $m$-D space can be split into $M$
dimensions, each of size 2, where:
\begin{equation}
    M =  \log s_0 + \log s_1 + \ldots + \log s_{m-1}.
\end{equation}

Similarly, the $n$-D index space can be split into $N$ dimensions, each of size 2, where:
\begin{equation}
    N =  \log s'_0 + \log s'_1 + \ldots + \log s'_{n-1}.
\end{equation}
Note that the dimension sizes in the co-ordinate and index spaces must be powers of two.
In this binary vector space representation, addition corresponds to XOR
operations and multiplication corresponds to AND operations, following the rules
of arithmetic in $F_2$.

\subsubsection{Basis Vector}
\label{sec:basis-vectors}

The linear layout is completely defined by specifying the index values for the
basis vectors of the binary coordinate space. For an $M$-D binary
coordinate space, the basis vectors are $(1, 0, 0, \ldots, 0), (0, 1, 0, \ldots,
0), \ldots, (0, 0, 0, \ldots, 1)$. For each basis vector $e_k$, the
corresponding index value $d_k$ in the $N$-D binary index space is
specified. The index value $(i_0, i_1, \ldots, i_{N-1})$ corresponding to any coordinate $(c_0, c_1, \ldots, c_{M-1})$
can then be computed using the following linear transformation where $d_k^j$
represents the $j$-th component of the index value $d_k$ corresponding to basis
vector $e_k$:
\begin{align}
    i_j &= d_0^j \cdot c_0 \oplus d_1^j \cdot c_1 \oplus \ldots \oplus d_{M-1}^j \cdot c_{M-1},
    \label{eq:linear-layout-transformation}
\end{align}
where, the $\cdot$ operator represents the AND operation and  the $\oplus$ operator
represents the XOR operation. Typically, to specify a linear layout, the basis
vectors and their corresponding index values are specified with respect to the
$m$-D coordinate space and $n$-D index space respectively.

\subsection{Polyhedral Model and Integer Set Library}
\label{sec:polyhedral-model-and-integer-set-library}

The polyhedral model provides a high-level mathematical representation of affine
loop-nests, capturing iteration spaces, array spaces, array accesses, and
schedules as integer polyhedra and relations between them. Advanced polyhedral
compilation techniques can then be employed to analyze and transform these
loop-nests for optimization.

In the polyhedral representation, each dynamic instance of a statement within a
loop-nest is mapped to an integer point in a bounded polyhedron. The
dimensionality of the loop-nest corresponds to the dimensionality of the
polyhedron, and the bounds on various loop indices correspond to the faces of
the polyhedron. This bounded polyhedron represents the iteration space of the
statement. Affine array accesses within statements can be represented as
mappings between iteration spaces and array spaces.

Polyhedral compilation techniques have proven powerful in optimizing loop-nests
in both domain-specific~\cite{baghdadi2019cgo, mullapudi2015asplos,
elango2018mapl, bhaskaracharya2020arxiv} and general-purpose~\cite{uday08pldi,
polly, verdoolaege2013taco, rstream} compilation contexts.

\subsubsection{Integer Set Library (ISL)}
\label{sec:isl}

The Integer Set Library (ISL)~\cite{isl} provides a comprehensive framework for
representing and manipulating integer sets and relations, making it an effective
tool for polyhedral modeling and transformation. ISL supports various
mathematical objects and operations that are fundamental to polyhedral
compilation.

\paragraph{Integer Sets}
An integer set in ISL represents a collection of integer points in a
multi-dimensional space. An integer set $S$ is defined as:
\begin{equation}
    S = \{ [x_0, x_1, \ldots, x_{n-1}] : \phi(x_0, x_1, \ldots, x_{n-1}) \},
\end{equation}
where $\phi$ is a conjunction of affine constraints involving the variables
$x_0, x_1, \ldots, x_{n-1}$. For example, the set $\{ [i, j] : 0 \leq i < N
\land 0 \leq j < M \}$ represents all integer points in an $N \times M$ region.

\paragraph{Integer Set Relations}
An integer set relation $R$ represents a mapping between two multi-dimensional
spaces. It is defined as:
\begin{equation}
    R = \{ [x_0, x_1, \ldots, x_{m-1}] \rightarrow [y_0, y_1, \ldots, y_{n-1}] : \psi(x_0, x_1, \ldots, x_{m-1}, y_0, y_1, \ldots, y_{n-1}  ) \},
\end{equation}
where $\psi$ is a conjunction of affine constraints involving input
and output variables $x_i$ and $y_j$. 

\paragraph{ISL Operations}
ISL provides a rich set of operations for manipulating integer sets and
relations.  Example operations on integer sets include union, intersection,
difference, and those on integer set relations include composition, inverse,
domain and range.

\paragraph{Quasi-Affine Extensions}

ISL extends beyond strictly affine constraints to support quasi-affine
expressions involving floor and ceiling operations. A quasi-affine constraint
has the following form where $a_i, b_j, c_{jk}, d_j, e_j, f$ are integer
constants:
\begin{equation}
    \sum_{i=0}^{n-1} a_i x_i + \sum_{j=0}^{m-1} b_j \lfloor \frac{\sum_{k=0}^{n-1} c_{jk} x_k + d_j}{e_j} \rfloor + f \geq 0.
\end{equation}
Essentially, in addition to being an affine expression involving $n$ variables,
the expression also involves an affine combination of $m$ floor division
operations, each of which involves the $n$ variables. This extension is crucial
for modeling complex loop transformations and layout mappings that involve
integer division operations.

\section{CuTe Layouts as Integer Set Relations}
\label{sec:cute}
In this section, we describe how the various mappings associated with a
given CuTe layout $H = s : d$, can be equivalently
expressed and manipulated as integer set relations using ISL.

\subsection{Coordinate and Index Spaces}
\label{coordinate-and-index-spaces}

Given a CuTe layout specification $H = s : d$ where $s = (s_0 , s_1 , \ldots{} ,
s_{n-1})$ and $d = (d_0 , d_1 , \ldots{} , d_{n-1})$, we first define the three
fundamental spaces that are essential for understanding the various mappings
specified by the layout.  The 1-D integral coordinate space $C$ is of size $P =
\prod_{i=0}^{n-1} s_i$, and it can be represented as the integer set:
\begin{equation}
\label{eq:integral-coordinate-space}
C = \{[c] : 0 \leq c < P\}.
\end{equation}
The $n$-D natural coordinate space $C_n$ is of shape $(s_0 , s_1 , \ldots{} ,
s_{n-1})$. So, it can be represented by the following $n$-D integer polyhedron:

\begin{align}
\label{eq:natural-coordinate-space}
C_n = &\{[c_0, c_1, \ldots, c_{n-1}] : 0 \leq c_0 < s_0 \land 0 \leq c_1 < s_1 \land \ldots{} \land 0 \leq c_{n-1} < s_{n-1}\}.
\end{align}

Note that while we use the term `spaces' for conceptual clarity, these are
technically represented as integer sets in ISL rather than ISL spaces. An ISL
space defines the dimensionality and naming of variables, while an integer set
represents a collection of integer points within a space. When we refer to
coordinate spaces or index spaces, we are referring to integer sets with
well-defined shapes or constraints that represent these conceptual spaces.

The exact form of the integer set representing the 1-D index space $I$ depends
on the layout $H$. In the following subsections, we shall see how
we can infer the integer set relations specifying the various mappings
associated with the layout $H$. 

\subsection{Coordinate Mapping}
\label{coordinate-mapping}

The coordinate mapping relates each point in the 1-D integral coordinate space
$C$ to a distinct point in the $n$-D natural coordinate space $C_n$. CuTe
layouts, by convention, employ a colexicographic ordering of the points in the
latter space. In order to obtain the integer set relation that represents this
mapping, consider the loop in Listing~\ref{lst:original-loop} whose iteration
space has the same size and shape as the coordinate space $C$. On tiling this
loop with tile size $s_0$, we get the nested loop in
Listing~\ref{lst:after-tiling-with-s0}.  Successively tiling the outer-most loop
with tile sizes $s_1 , s_2 , \ldots, s_{n-1}$ and reversing the loop order leads
to the transformed loop-nest shown in Listing~\ref{lst:after-successive-tiling}.

\begin{mdframed}[style=listingstyle]
\noindent\begin{minipage}[t]{0.5\textwidth}
\begin{lstlisting}[language=C, basicstyle=\ttfamily\footnotesize, keywordstyle=\color{codeblue}\bfseries, aboveskip=2pt, belowskip=0pt, escapeinside={(*@}{@*)}]
for (int (*@\textcolor{codepurple}{c}@*) = 0; (*@\textcolor{codepurple}{c}@*) < (*@\textcolor{codeorange}{P}@*); ++(*@\textcolor{codepurple}{c}@*))
  (*@\textcolor{codeblue}{\bfseries Stmt}@*)((*@\textcolor{codepurple}{c}@*));
\end{lstlisting}
\vspace{-0.8em}
\centering\small \captionof{lstlisting}{Original loop}\label{lst:original-loop}
\vspace{0.3em}
\begin{lstlisting}[language=C, basicstyle=\ttfamily\footnotesize, keywordstyle=\color{codeblue}\bfseries, aboveskip=2pt, belowskip=0pt, escapeinside={(*@}{@*)}]
for (int (*@\textcolor{codepurple}{cc}@*) = 0; (*@\textcolor{codepurple}{cc}@*) < (*@\textcolor{codeorange}{P}@*)/(*@\textcolor{codeorange}{s0}@*); ++(*@\textcolor{codepurple}{cc}@*))
  for (int (*@\textcolor{codepurple}{c0}@*) = 0; (*@\textcolor{codepurple}{c0}@*) < (*@\textcolor{codeorange}{s0}@*); ++(*@\textcolor{codepurple}{c0}@*))
    (*@\textcolor{codeblue}{\bfseries Stmt}@*)((*@\textcolor{codepurple}{cc}@*), (*@\textcolor{codepurple}{c0}@*));
\end{lstlisting}
\vspace{-0.8em}
\centering\small \captionof{lstlisting}{After tiling with $s_0$}\label{lst:after-tiling-with-s0}
\end{minipage}%
\hspace{4pt}
\begin{minipage}[t]{0.48\textwidth}
\begin{lstlisting}[language=C, basicstyle=\ttfamily\footnotesize, keywordstyle=\color{codeblue}\bfseries, aboveskip=2pt, belowskip=0pt, escapeinside={(*@}{@*)}]
for (int (*@\textcolor{codepurple}{c0}@*) = 0; (*@\textcolor{codepurple}{c0}@*) < (*@\textcolor{codeorange}{s0}@*); ++(*@\textcolor{codepurple}{c0}@*))
  for (int (*@\textcolor{codepurple}{c1}@*) = 0; (*@\textcolor{codepurple}{c1}@*) < (*@\textcolor{codeorange}{s1}@*); ++(*@\textcolor{codepurple}{c1}@*))
    ...
    for (int (*@\textcolor{codepurple}{cn\_1}@*) = 0; (*@\textcolor{codepurple}{cn\_1}@*) < (*@\textcolor{codeorange}{sn\_1}@*); ++(*@\textcolor{codepurple}{cn\_1}@*))
      (*@\textcolor{codeblue}{\bfseries Stmt}@*)((*@\textcolor{codepurple}{c0}@*), (*@\textcolor{codepurple}{c1}@*), ..., (*@\textcolor{codepurple}{cn\_1}@*));
\end{lstlisting}
\vspace{-0.8em}
\centering\small \captionof{lstlisting}{After successive tiling}\label{lst:after-successive-tiling}
\end{minipage}
\end{mdframed}

Essentially, the coordinate mapping $M_C^H$, can be understood as such a tiling
transformation. In the general form, $M_C^H$ can be captured by the following
integer set relation:
\begin{equation}
\label{eq:coordinate-mapping}
M_C^H = \{c \rightarrow [c \bmod s_0, \lfloor c/s_0 \rfloor \bmod s_1, \ldots, \lfloor c/\prod_{i=0}^{n-2} s_i \rfloor] : 0 \leq c < P\}
\end{equation}
Tiling is a quasi-affine transformation. So, effectively, non-trivial 1-D to
$n$-D coordinate mappings can be represented by quasi-affine integer set
relations. Note that the coordinate mapping $M_C^H$ depends only on the
shape $s$ of the layout. So, we will use the notation $M_C^s$ to represent the
coordinate mapping for a given shape $s$.

\subsection{Index Mapping}\label{index-mapping}

The index mapping $M_I^H$ of a layout $H = s : d$ maps the $n$-D natural
coordinate space $C_n$ to the 1-D index space $I$. By definition, this is the
dot product of the natural coordinate and the strides $d = (d_0 , d_1 ,
\ldots{} , d_{n-1}).$ Consequently, the index mapping $M_I^H$ is given by the
following affine relation:
\begin{equation}
\label{eq:index-mapping}
M_I^H = \{[c_0, \ldots, c_{n-1}] \rightarrow [\sum_{i=0}^{n-1} c_i d_i] : \forall i \in [0, n), 0 \leq c_i < s_i\}
\end{equation}
Intuitively, the index mapping $M_I^H$ is akin to specifying an array access
relation associated with the statement within the tiled loop-nest obtained above.
This is illustrated in the loop-nest shown in Listing~\ref{lst:tiled-loop-nest-with-access-expression}.
The access expression for a 1-D array space, in accordance with this index
mapping, would involve the loop iterators $c_0 , c_1 ,..., c_{n-1}$ with the
strides $d_0 , d_1 , \ldots{} , d_{n-1}$ serving as their co-efficients.  
\begin{mdframed}[style=listingstyle]
\begin{lstlisting}[
    language=C,
    basicstyle=\ttfamily\footnotesize,
    keywordstyle=\color{codeblue}\bfseries,
    commentstyle=\color{codegreen}\itshape,
    stringstyle=\color{codeorange},
    identifierstyle=\color{black},
    numbers=left,
    numberstyle=\tiny\color{codegray},
    numbersep=8pt,
    xleftmargin=12pt,
    frame=none,
    showstringspaces=false,
    tabsize=2,
    breaklines=true,
    escapeinside={(*@}{@*)}
]
for (int (*@\textcolor{codepurple}{c0}@*) = 0; (*@\textcolor{codepurple}{c0}@*) < (*@\textcolor{codeorange}{s0}@*); ++(*@\textcolor{codepurple}{c0}@*))
  for (int (*@\textcolor{codepurple}{c1}@*) = 0; (*@\textcolor{codepurple}{c1}@*) < (*@\textcolor{codeorange}{s1}@*); ++(*@\textcolor{codepurple}{c1}@*))
    ...
      for (int (*@\textcolor{codepurple}{cn\_1}@*) = 0; (*@\textcolor{codepurple}{cn\_1}@*) < (*@\textcolor{codeorange}{sn\_1}@*); ++(*@\textcolor{codepurple}{cn\_1}@*))
        (*@\textcolor{codeblue}{\bfseries Stmt}@*)((*@\textcolor{codepurple}{c0}@*), (*@\textcolor{codepurple}{c1}@*), ..., (*@\textcolor{codepurple}{cn\_1}@*), (*@\textcolor{codeorange}{I}@*)[(*@\textcolor{codepurple}{c0}@*)*(*@\textcolor{codeorange}{d0}@*) + (*@\textcolor{codepurple}{c1}@*)*(*@\textcolor{codeorange}{d1}@*) + ... + (*@\textcolor{codepurple}{cn\_1}@*)*(*@\textcolor{codeorange}{dn\_1}@*)]);
\end{lstlisting}
\captionof{lstlisting}{Tiled loop-nest with access expression}\label{lst:tiled-loop-nest-with-access-expression}
\end{mdframed}

\subsection{Layout Mapping}
\label{ssec:cute-layout-mapping}

The layout mapping $M_L^H$ associated with a layout $H$ is the mapping from the
1-D integral coordinate space to the 1-D index space. So, it is the composition
of the coordinate mapping and the index mapping. Overall, this approach to
obtain various mappings associated with a given layout $H$ is summarized in
Algorithm \ref{alg:layout-mapping}.
\begin{equation}
\label{eq:layout-mapping}
M_L^H = M_I^H \circ M_C^H
\end{equation}
\paragraph{Proof for correctness of Algorithm~\ref{alg:layout-mapping}:} Correctness follows by
construction: (1) \texttt{get\_coord\_mapping()} produces the unique tiling
transformation mapping 1-D integral coordinates to n-D natural coordinates via
colexicographic ordering (2) \texttt{get\_idx\_mapping()} constructs the affine
dot product relation mapping natural coordinates to indices using given strides
(3) ISL composition preserves relation semantics, giving the correct layout
mapping $M_L^H = M_I^H \circ M_C^H$.

Since the coordinate mapping $M_C^H$ is a quasi-affine relation, the layout
mapping $M_L^H$ is also a quasi-affine relation. The layout size $|H|$ is the
size of the layout mapping relation's domain i.e., its lexicographic maximum
plus 1. Likewise, the \textit{co-size} of the layout, represented as
$\left\|H\right\|$, is the lexicographic maximum of the layout mapping's range
plus 1. If the layout mapping relation $M_L^H$ is bijective, $H$ is considered
to be a \textit{bijective layout} only if its size and co-size are equal.
\begin{mdframed}[style=algorithmstyle]
\footnotesize
\begin{algorithm}[H]
\caption{Coordinate, Index and Layout Mapping for a given layout $H = s : d$}
\label{alg:layout-mapping}
\begin{algorithmic}[1]

\Require Layout $H = s : d$ where $s = (s_0, \ldots, s_{n-1}), d = (d_0, \ldots, d_{n-1})$
\Ensure Integer set relations for coordinate, index and layout mappings $M_C^H, M_I^H$ and $M_L^H$

\end{algorithmic}
\vspace{0.3em}
\noindent\begin{minipage}[t]{0.48\textwidth}
\begin{algorithmic}[1]
\Function{get\_layout\_mapping}{$s, d$}
    \State $c$ = get\_coord\_mapping$(s)$
    \State $i$ = get\_idx\_mapping$(s, d)$
    \State \Return $c.$apply\_range$(i)$ \Comment{ use ISL to compose}
\EndFunction

\State 

\Function{get\_idx\_mapping}{$s, d$}
    \State $index\_expr \gets [c_i \cdot d_i \mid i \in [0, \ldots, n-1]]$
    \State $bounds \gets [c_i \geq 0, c_i < s_i \mid i \in [0, \ldots, n-1]]$
    \State $dot\_prod \gets \sum index\_expr$
    \State $constraints \gets \bigwedge bounds$
    
    \State \Return $\{[c_0,\ldots,c_{n-1}] \rightarrow [dot\_prod] : constraints\}$
\EndFunction
\end{algorithmic}
\end{minipage}%
\hfill
\begin{minipage}[t]{0.48\textwidth}
\begin{algorithmic}[1]
\Function{get\_coord\_mapping}{$s$}
    \State $strides \gets (1, s_0, s_0s_1, \ldots{}, s_0s_1\ldots{}s_{n-2})$
    \State $in\_bounds \gets [c \geq 0, c < \prod_{i=0}^{n-1} s_i]$
    \State $out\_bounds \gets \bigwedge [c_i \geq 0, c_i < s_i \mid i \in [0,\ldots,n-1]]$
    \State $constraints \gets in\_bounds \land out\_bounds$
    
    \For{$i \in [n-1,\ldots,0]$}
        \State $cnst \gets \emptyset$
        \If{$i = n-1$}
            \State $cnst \gets c_{n-1} = \lfloor c/strides_{n-1} \rfloor$
        \Else
            \State $cnst \gets c_i = \lfloor (c \bmod strides_{i+1})/strides_i \rfloor$
        \EndIf
        \State $constraints \gets constraints \land cnst$
    \EndFor
    
    \State \Return $\{[c] \rightarrow [c_0,\ldots,c_{n-1}] : constraints\}$
\EndFunction
\end{algorithmic}
\end{minipage}
\end{algorithm}
\normalsize
\end{mdframed}
Note that these mappings could also be used to express a schedule
transformation, in which case the index space essentially corresponds to a
transformed iteration space. For example, the colexicographic ordering that is
used to obtain the coordinate mapping implies that the colexicographic strides
$(1, s_0, s_0s_1, \ldots{} , s_0s_1\ldots s_{n-2})$ can be used to linearize the
natural coordinates (or iteration space of the tiled loop-nest) back to the
integral coordinates (or the original 1-D iteration space). So, if the strides
$d_0, d_1, \ldots, d_{n-1}$ are the same as these colexicographic strides, the
index mapping $M_I^H$ essentially represents such a linearization. In loop
transformation terms, effectively, this means that the overall layout mapping
$M_L^H$ represents an identity schedule. On the other hand, suppose the layout
is bijective but with strides $d$ that are not the same as the colexicographic
strides. In this case, the index mapping $M_I^H$ is essentially like a schedule
mapping that involves a permutation of the dimensions in the natural coordinate
space (or the loops in the tiled loop-nest) followed by a colexicographic
linearization of this permuted schedule.

\subsection{Hierarchical Layouts}
\label{hierarchical-layouts}

So far, in this section, we have focused on expressing non-hierarchical CuTe
layouts as integer set relations. However, recall that CuTe also supports
hierarchical layouts with nested shapes and strides e.g. $H = (4, (2, 2)) : (2,
(1, 8))$. We now describe how the same approach can be extended to model
hierarchical layouts as well.

Suppose $H = s : d$ is a hierarchical layout with some arbitrary
nesting of shapes and strides. A layout $H_f = s_f : d_f$ can then be inferred
by flattening the nested shape and stride tuples $s$ and $d$. The layout mapping
$M_{L}^{H_f}$ for the flattened layout can then be computed using Algorithm
\ref{alg:layout-mapping}. Note that the layout mapping $M_L^{H}$ for the
hierarchical layout is the same as $M_{L}^{H_f}$.

Hierarchical layouts define multiple different natural coordinate spaces for
the same layout specification $H = s : d$, with the overall layout mapping being
the same. Suppose $s'$ is the shape of one such coordinate space. The size of
this coordinate space, i.e., the number of integer points in it, must be the
same as that of the flattened shape $s_f$, even though the rank of $s'$ must be
less than that of $s_f$.  The coordinate mapping $M_{C}^{s'}$ for the shape
$s'$ can be computed using the \small\texttt{get\_coord\_mapping()}\normalsize\
function described in Algorithm \ref{alg:layout-mapping}. Given the layout
mapping $M_{L}^{H_f}$ and the coordinate mapping $M_{C}^{s'}$ for the shape
$s'$, the corresponding index mapping $M_{I}^{s'}$ is given by the composition
$M_{L}^{H_f} \circ (M_{C}^{s'})^{-1}$ where $(M_{C}^{s'})^{-1}$ represents the
inverse of $M_{C}^{s'}.$ In this way, it is possible to compute the associated
coordinate and index mappings for every natural coordinate space supported by
a hierarchical layout. However, the index mappings computed for such shapes are
not necessarily simple dot products and could be quasi-affine.  For example,
consider the hierarchical layout $H = (4, (2, 2)):(2, (1, 8))$. To compute the
layout mapping, we consider the flattened layout $H_f = (4, 2, 2) : (2, 1, 8)$.
The resulting layout mapping is given by:
\begin{align}
    M_{L}^{H_f} = &\{[c] \rightarrow [7 + 2c + 6\lfloor c/8 \rfloor + 7\lfloor (-1 - c)/4 \rfloor] : 0 \leq c \leq 15\}. \notag
\end{align}
The corresponding coordinate and index mappings are given by:
\begin{align}
    M_{C}^{H_f} = &\{[c] \rightarrow [c \bmod 4, \lfloor c/4 \rfloor - 2\lfloor c/8 \rfloor, \lfloor c/8 \rfloor] : 0 \leq c \leq 15\} \notag \\
    M_{I}^{H_f} = &\{[c_0, c_1, c_2] \rightarrow [2c_0 + c_1 + 8c_2] : 0 \leq c_0 \leq 3 \land 0 \leq c_1 \leq 1 \land 0 \leq c_2 \leq 1\}. \notag
\end{align}
 Note that the
index mapping is a dot product relation as expected.
Now, the shape $(4, 4)$ is clearly one of the shapes supported by the hierarchical
layout. The associated coordinate mapping $M_{C}^{(4,4)}$ is given by:
\begin{align}
    M_{C}^{(4,4)} = &\{[c] \rightarrow [c \bmod 4, \lfloor c/4 \rfloor] : 0 \leq c \leq 15\}. \notag
\end{align}
Finally, the corresponding index mapping for this 2-D coordinate space
can be computed as $M_{L}^{H_f} \circ (M_{C}^{(4,4)})^{-1}$ which evaluates to the
following quasi-affine relation:
\begin{align}
    M_{I}^{(4,4)} = &\{[c_0, c_1] \rightarrow [(-3 + 2c_0 + 4c_1 + 3((1 + c_1) \bmod 2))] : 0 \leq c_0 \leq 3 \land 0 \leq c_1 \leq 3\}. \notag
\end{align}
\subsection{CuTe Layout From Layout Mappings And Shape}
\label{cute-layout-from-layout-mappings-and-shape}

In the previous subsections, we have seen how a layout mapping can be
represented as an integer set relation given a CuTe layout specification $H = s
: d.$ We now consider the inverse problem of inferring a CuTe layout
specification from its layout mapping, specified as an integer set relation,
given its shape.

Consider the scenario when the layout mapping $M_L^H$ of a layout $H$ and only
its shape $s$ are known. If the shape $s$ is nested, it can be flattened and
once the strides are found for the flattened shape, the nesting can be
introduced back into the shape and stride tuples as required. This
reconstruction is deterministic because hierarchical layouts define multiple
coordinate spaces for the same layout mapping. Reconstruction simply re-nests
the flattened stride tuple to match the original nesting pattern. So, for the
sake of simplicity, we assume that the layout is not hierarchical so that that
shape $s$ is not nested.

The shape alone is enough to derive the coordinate mapping $M_C^H$
using the \small\texttt{get\_coord\_mapping()}\normalsize\ function described
in Algorithm \ref{alg:layout-mapping}. Furthermore, the index mapping $M_I^H$
can be computed as $M_L^H \circ (M_C^H)^{-1}$. If the index mapping is a strictly affine
relation, the co-efficients of its input dimensions give the strides $d$ required
to completely define the layout $H$. This approach is summarized in
Algorithm \ref{alg:get-layout-strictly-affine}.
\footnotesize
\begin{mdframed}[style=algorithmstyle]
\begin{algorithm}[H]
\caption{Construct Layout from Strictly Affine Index Mapping and Shape}
\label{alg:get-layout-strictly-affine}
\begin{algorithmic}[1]

\Require $i$: integer set relation representing index mapping, $s$: shape tuple of layout (can be nested)
\Ensure Resulting layout $H$

\Function{get\_layout\_strictly\_affine}{$i, s$}
    \State $assert$(i is strictly affine) 
    \State $eq\_cnsts \gets \text{get\_eq\_constraints\_involving\_input\_dims}(i)$ \Comment{get equality constraints}
    \State $assert$(len(eq\_cnsts) == 1) \Comment{there must be exactly one equality constraint}
    \State $eq\_cnst \gets eq\_cnsts[0]$ \Comment{get the single equality constraint}
    \State $n\_dims \gets  i$.dim(isl\_dim\_type.in\_) \Comment{get the number of input dimensions}
    \State $d \gets [eq\_cnst.$\text{get\_coeff}(i) for\ $i$\ in\ \text{range}($n\_dims$)$]$ \Comment{get the strides}
    \State \Return $H = (s : \text{unflatten}(d,s))$ \Comment{strides must be nested similar to the shape}
\EndFunction

\end{algorithmic}
\end{algorithm}
\end{mdframed}
\normalsize
\paragraph{Proof for correctness of Algorithm~\ref{alg:get-layout-strictly-affine}:} A strictly affine index
mapping has exactly one equality constraint of the form $o = \sum_j d_j c_j$
where $o$ is the output and $c_j$ are inputs.  The coefficients $d_j$ are by
definition the strides. We thus get the unique layout $(s : d)$
whose index mapping equals the input relation.

On the other hand, if the index mapping is not a strictly affine relation, no
straightforward layout representation exists for the given shape although
another shape with the same size could have a valid CuTe layout representation
that leads to the given layout mapping. For example, suppose the layout mapping
$M_L^H$ was obtained for the layout $H = (4, (2, 2)) : (2, (1, 8))$. But as we
have seen in the section, if it is then used to determine a valid CuTe layout
for the shape $(4, 4)$, the index mapping inferred will be quasi-affine. If the
shape $(4, 2, 2)$ had been used instead, the index mapping inferred would be
strictly affine.

\subsection{CuTe Layout From Layout Mappings and Strides}
\label{strides}

Consider the scenario when the layout mapping $M_L^H$ and the strides $d$ are
known, but not the shape $s$. Again, without loss of generality, we assume here
that the layout is not hierarchical. 

The rank $n$ of the strides tuple is essentially the dimensionality of the
natural coordinate space. Furthermore, note that the lexicographic maximum of
the range of the layout mapping relation $M_L^H$ is $\left\|H\right\|-1$ i.e.,
the co-size of the layout minus 1. Consequently, the strides can be used to
define an $n$-D coordinate to 1-D index dot product relation with the strides
serving as the co-efficients.  The only bounds on the dimensions involved in
this relation are the following: (a) the coordinate space dimensions are
non-negative and (b) the single output dimension in the index space is fixed to
be equal to $\left\|H\right\|- 1.$

Essentially, there must exist an integer point $p$ in the domain of this
relation such that, if $s = p + 1$ is the shape $s$ obtained from component-wise
addition, the inferred layout $s : d$ has a layout mapping that equals $M_L^H$.
Every point $p$ in the domain of this relation represents a non-negative
solution to the linear equation $ d_0x_0 + d_1x_1 + \ldots + d_{n-1}x_{n-1} =
\left\|H\right\|- 1$ and exactly one of them must correspond to the
lexicographic maximum of the coordinate space for the layout $H$. If there is
no such point, no corresponding layout can be found for the layout mapping
$M_L^H$ and the given strides $d$. This approach is summarized in Algorithm
\ref{alg:get-layout-from-layout-mapping-and-strides}.

There are many libraries that can be used for efficiently solving such
linear equations. ISL itself comes with its own solver. However, for the
sake of efficiency, we employed PuLP, a python library for modeling
linear and mixed integer programming problems, as it allows us to employ
various different solvers such as CBC, GLPK, CPLEX etc.
\\
\\
Overall, in order to infer a CuTe layout from its layout mapping, either the
shape or the strides of the layout must be known, in addition to the integer set
relation for the layout mapping. The problem of inferring both the shape and the
strides from the layout mapping alone is still an open one.  In loop
transformation terms, this is equivalent to determining a tiling transformation
for a 1-D iteration space that would turn a 1-D to 1-D quasi-affine array access
relation associated with it into an n-D to 1-D strictly affine array access
relation.

\footnotesize
\begin{mdframed}[style=algorithmstyle]
\begin{algorithm}[H]
\caption{Construct Layout from Layout Mapping and Strides}
\label{alg:get-layout-from-layout-mapping-and-strides}
\begin{algorithmic}[1]

\Require $l$: integer set relation representing layout mapping, $d$: strides of required layout
\Ensure Resulting layout $H$ or $\text{None}$ if no valid layout found

\Function{get\_layout\_from\_strides}{$l, d$}
    \State $max\_index \gets \text{maximum value in range of } l$ 
    \State $points \gets \text{find\_non\_negative\_solutions}(d, max\_index)$ \Comment{find all non-negative solutions to the linear equation $d_0x_0 + d_1x_1 + \ldots + d_{n-1}x_{n-1} = max\_index$}    
    
    \For{each $p \in points$}
        \State $s \gets p + 1$ \Comment{Add 1 to each component}
        \State $H \gets (s : d)$
        \If{$get\_layout\_mapping(s, d) = l$} 
            \State \Return $H$
        \EndIf
    \EndFor
    
    \State \Return $\text{None}$ \Comment{No valid layout found}
\EndFunction

\end{algorithmic}
\end{algorithm}
\end{mdframed}
\normalsize

\paragraph{Proof for correctness of Algorithm~\ref{alg:get-layout-from-layout-mapping-and-strides}:}
By definition, the maximum index value in $l$ occurs at the lexicographic
maximum of the coordinate space. For layout $(s:d)$, this maximum coordinate is
$(s_0-1, \ldots, s_{n-1}-1)$, giving index $\sum_i d_i(s_i-1)$.  Setting $p_i =
s_i - 1$, we get $\sum_i d_i p_i = \text{max\_index}$. The algorithm
exhaustively checks all non-negative solutions, ensuring correctness by
verification.

\subsection{Layout Operations}
\label{layout-operations}

CuTe provides a rich set of operations that transform layouts into new layouts.
The integer set relation approach described in previous subsections applies to
any layout, including those resulting from CuTe operations.
This section demonstrates how three fundamental operations -- composition, inverse, and
complement -- can be expressed and computed using ISL integer set operations.
Other CuTe operations such as layout divisions and products are built upon these
three fundamental operations.

\subsubsection{Composition}
\label{composition}

A layout composition, $H = G \circ F,$ defines a layout mapping from
the coordinate space of $F$ to a 1-D index space. In this
subsection, we discuss how $H$ can be inferred using the integer
set relations associated with the layouts $G$ and $F$.

\paragraph{Size of $G$ at least as large as co-size of $F$}
\label{size-of-g-at-least-as-large-as-co-size-of-f}
Generally, the resulting layout mapping $M_L^H$ is the same as the composition
of the layout mappings for the layouts $G$ and $F$ i.e., $M_L^G \circ M_L^F$.
However, this is not true when the latter has holes in its domain. To better
understand this, consider the example layout composition, $G \circ F = (2, 1) :
(1, 80) \circ (2, 2) : (2, 1).$ The resulting CuTe layout is $H = (2, 2) : (80,
1)$.  Note that the shape of $H$ is same as that of $F$. This follows from the
definition of the composition operation in CuTe which mandates that the
resulting layout $H$ must have a shape that is the same, or is compatible with
that of $F.$ On the other hand, if we compose the integer set relations $M_L^G$
and $M_L^F$ using ISL, it can be seen below that ISL restricts the domain of the
composition to include just 0 and 2.  This is in accordance with standard rules
of relation composition.
\begin{align}
    M_L^F = &\{ [c] \rightarrow [-c + 3\lfloor(1 + c)/2\rfloor] : 0 \leq c \leq 3 \} \notag \\
    M_L^G = &\{ [c] \rightarrow [c] : 0 \leq c \leq 1 \} \notag \\
    M_L^G \circ M_L^F = &\{ [c] \rightarrow [\lfloor c/2 \rfloor] : (c) \bmod 2 = 0 \land 0 \leq c \leq 2 \} \notag
\end{align} 
The difference is mainly because CuTe implicitly `promotes' $G$ to a layout $G'$
such that the size of the layout $G'$ equals the co-size of the layout $F$. This
promotion involves increasing the shape only along the last dimension. So, in
this case, $G= (2, 1) : (1, 80)$ gets promoted to $G' = (2, 2) : (1, 80)$ and
the actual composition that is computed by CuTe is $G' \circ F$. Consequently, a
straightforward composition of integer relations $M_G \circ M_F$ does not
necessarily equal the layout mapping of the composed layout $H$ when there are
holes in the domain of $M_G \circ M_F$.

This deviation from standard relation composition semantics has critical
implications for memory safety and program correctness. When CuTe performs
implicit layout promotion, the composed layout produces mappings that extend
beyond what relational composition semantics would yield. For instance, in the
example above, $F$ contains mappings $\{(0,0), (1,2), (2,1), (3,3)\}$ while $G$
contains only $\{(0,0), (1,1)\}$. Relational composition should yield mappings
$\{(0,0), (2,1)\}$. Yet CuTe's composed layout produces $\{(0,0), (1,80), (2,1),
(3,81)\}$. Code generated from such compositions without proper bound checking
can lead to out-of-bound memory accesses. Given the lack of precise modeling of
constraints on the range of layout mappings in CuTe, the onus is on the
programmer to track and prevent such errors.  However, with integer set
relations, the required predicates can be automatically synthesized.

So, for the rest of this section, we
assume that there are no holes in the domain of $M_G \circ M_F$ i.e, no implicit
layout promotion is necessary as size of $G$ is at least as large as the co-size
of $F$.

\paragraph{By-Mode Composition}
\label{by-mode-composition}

CuTe supports by-mode composition, which decomposes layout composition into
separate computations for each dimension or mode of the input layout, where the
number of modes equals the rank of the layout. So, if a layout $F$ can be
expressed as a concatenation of rank-1 layouts $(F_0, F_1, \ldots, F_{n-1})$,
then the composition $G \circ F$ can be computed as $G \circ F = (G \circ F_0, G
\circ F_1, \ldots, G \circ F_{n-1})$. For example, given $F = (2, 2) : (2, 1)$
and $G = (2, 2) : (1, 80)$, the layout $F$ decomposes into $F_0 = 2 : 2$ and
$F_1 = 2 : 1$, yielding $G \circ F_0 = 2 : 80$ and $G \circ F_1 = 2 : 1$, which when
concatenated gives $G \circ F = (2, 2) : (80, 1)$. For the
remainder of this section, we assume without loss of generality that $F$ is a
rank-1 layout, which simplifies the analysis while preserving the essential
concepts.

\paragraph{Composition with 1-D Layout}
\label{composition-with-1-d-layout}
Now, given the above assumptions, if we consider the natural coordinate space
of the composed layout $H = G \circ F$ where $F$ is a layout of rank 1, either
of the following two scenarios are possible:

\begin{itemize}
\item
  $H$ has the same 1-D natural coordinate space as the layout $F$ in
  which case the index mapping must be strictly affine
\item
  $H$ has the same integral coordinate space but the natural coordinate space
  is multi-dimensional i.e., the shape of $H$ is compatible with that of $F$ but
  not the same. In this case, a mapping from the integral coordinate space of
  $F$ to the natural coordinate space of $G$ can be inferred from the layout
  mapping $M_L^F$ and the coordinate mapping $M_C^G$. The range of this mapping
  must be such that all dimensions have constant strides and bounds. Using these
  strides and bounds, the shape of $H = G \circ F$ can be
  computed. Finally, given the shape and the overall layout mapping $M_L^G \circ
  M_L^F$ , the required layout $H$ can be inferred as discussed in
  Section~\ref{cute-layout-from-layout-mappings-and-shape}.
\end{itemize}
\begin{mdframed}[style=algorithmstyle]
\footnotesize
\begin{algorithm}[H]
\caption{Layout Composition}
\label{alg:layout-composition}
\begin{algorithmic}[1]

\Require Two layouts $g$ and $f$
\Ensure Composed layout

\Function{compose}{$g, f$}
    \If{$f.rank = 1$}
        \State $gl \gets $get\_layout\_mapping$(g.shape, g.strides)$
        \State $gc \gets $get\_coord\_mapping$(g.shape)$
        \State $fl \gets $get\_layout\_mapping$(f.shape, f.strides)$
        \State $fc \gets $get\_coord\_mapping$(f.shape)$
        
        \State $l \gets fl.$apply\_range$(gl)$ \Comment{get composed layout mapping}
        \State $i \gets fc.$reverse$().$apply\_range$(l)$ \Comment{get index mapping for 1-D shape}
        
        \If{$i$ is strictly affine} \Comment{natural coordinates of composed layout same as those of $f$}
            \State \Return \text{get\_layout\_strictly\_affine}$(i, f.shape)$ \Comment{Algorithm~\ref{alg:get-layout-strictly-affine}}
        \Else \Comment{natural coordinate space of composed layout is multi-dimensional}
            \State $r \gets fl.$apply\_range$(gc).$range$()$  
            \State $s \gets $\text{extract\_shape}$(r)$ \Comment{r has constant strides and bounds for each dim, else invalid composition}
            \State $s\gets \langle x \mid x \in s, x \neq 1 \rangle$ \Comment {drop unit dims}
            \State $c \gets $\text{get\_coord\_mapping}$(s)$
            \State $i \gets c.$reverse$().$apply\_range$(l)$
            \State \Return \text{get\_layout\_strictly\_affine}$(i, shape)$
        \EndIf
    \Else \Comment{employ by-mode composition}
        \State $f_0 \gets \text{Layout}(f.shape[0], f.strides[0])$
        \State $res \gets g.$compose$(f_0)$
        \State $layouts \gets \emptyset$
        
        \For{$i \in [1..\text{len}(f.shape)-1]$}
            \State $f_i \gets \text{Layout}$$(f.shape[i], f.strides[i])$
            \State Append $g.$compose$(f_i)$ to $layouts$
        \EndFor

        \State \Return $res.$concat$(layouts)$ if $layouts \neq \emptyset$ else $res$
    \EndIf
\EndFunction

\end{algorithmic}
\end{algorithm}
\normalsize
\end{mdframed}
This approach for computing layout composition using the various integer set
relations associated with the input layouts is summarized in
Algorithm~\ref{alg:layout-composition}. Note that it employs the function
\small\texttt{extract\_shape()}\normalsize~to determine the shape of the resulting layout given an
integer set. Essentially, this is akin to determining the shape of a perfect
loop-nest given an integer set specification of the iteration space, assuming
zero-based indexing with unit strides and constant upper bound for each loop
iterator. In some cases, when such a loop-nest cannot be inferred, then the
layout composition can be considered invalid.
\paragraph{Proof for correctness of Algorithm~\ref{alg:layout-composition}:} For rank-1 layouts, the composed mapping $l = M_L^G \circ M_L^F$
correctly represents $G \circ F$ by relation composition semantics. When the
resulting index mapping is strictly affine, the natural coordinate space equals
that of $F$. Otherwise, the shape is extracted from the range of $M_C^G \circ
M_L^F$, giving the correct multi-dimensional coordinate space. For higher-rank
layouts, by-mode composition preserves CuTe semantics by decomposing into rank-1
subproblems and concatenating results. This is a property of CuTe layouts as
described in the CuTe documentation~\cite{cute}.

Consider the example of layout composition with $G = (2, 2) : (1, 80)$ and $F =
(2, 2) : (2, 1)$.  With by-mode composition, we compute $G \circ F_0 = (2, 2) :
(1, 80) \circ 2 : 2$ and $G \circ F_1 = (2, 2) : (1, 80) \circ 2 : 1$
separately. The layout mapping $M_L^{G \circ F_0}$ is the following strictly
affine integer set relation:
\begin{align}
M_L^{G \circ F_0} = \{ [c] \rightarrow [80c] : 0 \leq c \leq 1 \}. \notag
\end{align}
This implies that the shape of the composed layout $G \circ F_0$ is 2, same as
the shape of $F_0$. Furthemore, the coefficient 80 in the above relation implies
that the stride of $G \circ F_0$ is 80 giving us the layout $G \circ F_0 = 2 :
80$. Similarly, for $G \circ F_1$, the layout mapping $M_L^{G \circ F_1}$ is
as follows:
\begin{align}
M_L^{G \circ F_1} = \{ [c] \rightarrow [c] : 0 \leq c \leq 1 \}. \notag
\end{align}
This implies that $G \circ F_1 = 2: 1$. Concatenating the two layouts, we
get the composed layout $G \circ F = (2, 2) : (80, 1)$.

Now, consider another commposition $G \circ F = (4, 6, 8, 10) : (2, 3, 5, 7) \circ 6 : 12$.
The layout mapping $M_L^{G \circ F}$ is the following quasi-affine integer set relation:
\begin{align}
M_L^{G \circ F} = \{ [c] \rightarrow [(-4c + 13\lfloor((1 + c)/2)\rfloor)] : 0 \leq c \leq 5 \}. \notag
\end{align}
This means that the shape of the composed layout $G \circ F$ is not the same as
that of $F$. The range of the mapping from the integral coordinate space of $F$
to the natural coordinate space of $G$ is given by the following integer set:
\begin{align}
S &= \{ [c_0, c_1, c_2, c_3] : c_3 = 0 \land c_0 = 0 \land c_2 \geq 0 \land 2 - c_2 \geq 0 \notag \\
&\land -c_1 + 3\lfloor((c_1)/3)\rfloor = 0 \land 3 - c_1 \geq 0 \land c_1 \geq 0 \}. \notag
\end{align}
The constraints on $c_0$ and $c_3$ imply that both are fixed to 0. From the
constraints on $c_1$, we can infer that $c_1$ is a multiple of 3 and is in the
range $[0, 3]$ and the constraints on $c_2$ imply that it has a unit stride with
the range $[0, 2]$. So, the shape of this space turns out to be $(1, \lceil (3 -
0 + 1) / 1 \rceil, 3, 1)$ i.e., $(2, 3)$ after droppping the unit size
dimensions. The composition of the inverse of the coordinate mapping for the shape $(2, 3)$
and the layout mapping $M_L^{G \circ F}$ gives us the following index mapping:
\begin{align}
M_I^{G \circ F} = \{ [c_0, c_1] \rightarrow [(9c_0 + 5c_1)] : 0 \leq c_0 \leq 1 \land 0 \leq c_1 \leq 2 \}. \notag
\end{align}
This implies that the strides of the composed layout $G \circ F$ are $(9, 5)$
giving us the layout $G \circ F = (2, 3) : (9, 5)$.

\subsubsection{Complement}
\label{complement}

CuTe supports a complement operation that takes a layout $H$ and a target size
$D$ to compute a layout $H^\prime$ that fills up the `gaps' in the index space
of $H$, up to size $D$. Basically, if we consider the layout mapping relation
$M_L^{H}$ for $H$, the range of this relation may not be a single convex integer
set and so, may have gaps. Recall from Section~\ref{sec:layout-operations} that
the layout complement $H^\prime$ is such that if $HH^\prime$ is the
concatenation of $H$ and $H^\prime$, the layout mappping relation
$M_L^{HH^\prime}$ fills up as many gaps as possible in the interval $[0, D)$.  


Consider the set of integers $Z_D$ in the interval $[0, D)$. The problem of
determining the complement layout $H^\prime$ can be seen as one of trying to map
to as many points in this set as possible. This can be done by building up
another layout $H_\mathit{curr}$, starting with the layout $H$, and then
iteratively concatenating more layouts to it to eventually find the layout
$H_\mathit{curr} = HH^\prime$. In order to do this, we need to iteratively (a)
find the lexicographically smallest gap (in terms of where the gap begins) in
the index space of $H_\mathit{curr}$ and (b) fill it up by repeating the mapping
of $H_{\mathit{curr}}$ a certain number of times and (c) update $H_\mathit{curr}$ to include the
newly filled gap. Finally, if $D$ exceeds the maximum index that is mapped to by
$H$, we need to similarly fill up the gap represented by the set of integers in
the interval $[\left\|H\right\|, D)$.  Algorithm~\ref{alg:layout-complement}
describes such an approach. 

It keeps track of the number of integers that are considered as $\mathit{filled}$ with
$H^\prime$ initialized to $1:0$ so that initially, $\mathit{filled} = 1$. The beginning point,
$\mathit{begin}$, of the gap to be filled in the index space of $H_\mathit{curr}$ is found by
subtracting $Z_{\mathit{filled}}$ as well as the range of $H_{\mathit{curr}}$ from $Z_D$, and
then finding the lexicographically smallest integer in this set difference.  If
this point lies in the index space of $H$, then the end point, $\mathit{end}$, of the gap
(treating the gap as a half-open interval that does not include the end point)
is similarly found as the lexicographically smallest integer in the set
difference of the range of $H$ and $Z_{\mathit{filled}}$, after updating
$\mathit{filled}$ to cover the interval up to the beginning point. This gap is
then filled up by concatenating a layout of shape $\lfloor \mathit{end}/\mathit{begin} \rfloor$
and stride $\mathit{begin}$ to $H_{\mathit{curr}}$.  If this shape is 0, then we
ignore such layouts as the gap cannot really be filled by a valid layout
mapping and so, the algorithm continues updating $\mathit{filled}$ to skip this gap.
Finally, if there is any gap beyond the index space of $H$ (due to $D \geq \left\|H\right\|$)
then the algorithm repeats the mapping of $H$ to fill up the interval $[\left\|H\right\|, D)$.

Consider the layout $H = (2, 2) : (1, 5)$ and suppose we want to compute its
complement with target size $D = 20$. We start with $H_{curr} = H$ so that the
range of the layout mapping $M_L^{H_{curr}}$ is given by the following integer set:
\begin{align}
S = \{ [c] : -3 \leq c \leq 6 \land 5\lfloor((-2 + c)/5)\rfloor \leq -5 + c \land 5\lfloor((-1 - 2c)/5)\rfloor \geq -6 - 3c \}. \notag 
\end{align}
\footnotesize
\begin{mdframed}[style=algorithmstyle]
\begin{algorithm}[H]
\caption{Layout Complement}
\label{alg:layout-complement}
\begin{algorithmic}[1]
\Require Tensor layout $H$, size $D$
\Ensure Complement layout $H^\prime$ 

\State $\mathit{max\_index} \gets \max(D-1, \text{cosize}(H)-1)$ \Comment{max index upto which to fill}
\State $\mathit{filled} \gets [0,1)$ \Comment{Initial filled interval}
\State $H^\prime \gets Layout(1, 0)$ \Comment{Trivial complement layout}
\State $H_{\mathit{curr}} \gets H$
\State $\mathit{end} \gets 0$

\While{$\mathit{end} \leq \mathit{max\_index}$} \Comment{fill gaps until max index is reached}
    \State $\mathit{gaps} \gets [0,\mathit{max\_index}+1)$ \Comment{all possible gaps}
    \State $gaps \gets gaps \setminus filled$ \Comment{remove already filled gaps}
    \State $h \gets $get\_layout\_mapping$(H_{\mathit{curr}}.shape, H_{\mathit{curr}}.strides)$ 
    \State $\mathit{gaps} \gets \mathit{gaps} \setminus h.range()$ \Comment{remove gaps that are already mapped}

    \If{$\mathit{gaps} = \emptyset$}  \Comment{no more gaps to fill}
         \State \textbf{break} 
    \EndIf 
    
    \State $\mathit{begin} \gets lexmin(\mathit{gaps})$
    \State $\mathit{filled} \gets [0,\mathit{begin})$
    
    \If{$\mathit{begin} < \text{cosize}(H)$} \Comment{gap falls in range of layout $H$}
        \State $\mathit{end} \gets lexmin(\text{range}(H) \setminus \mathit{filled})$
        \State $\mathit{shape}, \mathit{stride} \gets \lfloor \mathit{end}/\mathit{begin} \rfloor, \mathit{begin}$
    \Else \Comment{gap falls outside range of layout $H$}
        \State $\mathit{begin}, \mathit{end} \gets \text{cosize}(H), \mathit{max\_index} + 1$
        \State $\mathit{shape}, \mathit{stride} \gets \lceil \mathit{end}/\mathit{begin} \rceil, \mathit{begin}$
    \EndIf
    
    \If{$shape > 1$} \Comment{gap can be filled by repeating layout $H_{curr}$}
        \State $c = Layout(shape, stride)$ \Comment{rank-1 layout to repeat $H_{curr}$}
        \State $H_{\mathit{curr}} \gets H_{\mathit{curr}}.concat(c)$ \Comment{concatenate to current layout}
        \If {$H^\prime$ is a trivial complement} 
             \State $H^\prime \gets H^\prime.concat(c)$
        \Else
            \State $H^\prime \gets c$
        \EndIf
        
    \EndIf
    
    \State $\mathit{filled} \gets [0,\mathit{end})$
\EndWhile

\State \Return $H^\prime$
\end{algorithmic}
\end{algorithm}
\end{mdframed}
\normalsize
Subtrating this range from the set of integers in $[0, 20)$ gives us the
following integer set:
\begin{align}
   R = \{ [c] : 0 < c \leq 19 \land (c \geq 7 \lor (c \leq 6 \land 5\lfloor((-2 + c)/5)\rfloor \geq -4 + c)) \}. \notag
\end{align}
The lexicographically smallest integer in this set is $2$. Subtracting the set
of integers in the interval $[0, 2)$ from set S gives us an integer set whose
lexmin is $5$. Consequently, the first gap to be filled is represented by the
set of integers in the interval $[2, 5)$. This can be filled by repeating the
layout $H$ $\lfloor(5/2)\rfloor = 2$ times with a stride of 2.

After updating $H_{curr}$ to be equal to $(2, 2, 2) : (1, 5, 2)$, we repeat this
process to find the next gap. The beginning of the gap is found to be $9$. As
this falls outside the range of $H$, the gap to be filled is given by the set of
integers in the interval $[9, 20)$. This can be filled by repeating the layout
$H$ $\lceil(20/9)\rceil = 3$ times with a stride of 9. Overall, the complement
layout $H^\prime$ is given by the following layout $H^\prime = (2, 3) : (2, 9)$.

\paragraph{Proof for correctness of Algorithm~\ref{alg:layout-complement}:} The algorithm maintains
the invariant that `filled' represents indices covered by $H_{\text{curr}}$. In
each iteration, it finds the lexicographically smallest gap and fills it by
repeating $H_{\text{curr}}$ with appropriate stride and shape. The shape
calculation $\lfloor \text{end}/\text{begin} \rfloor$ for internal gaps and
$\lceil \text{end}/\text{begin} \rceil$ for trailing gaps ensures optimal
coverage. Termination is guaranteed as `end' monotonically increases towards `max\_index'.

\subsubsection{Inverse}
\label{inverse}

Consider a layout $H = s : d$ that is bijective. As described in
Sections~\ref{coordinate-mapping} and~\ref{ssec:cute-layout-mapping}, the coordinate
mapping can be seen as a tiling transformation. Furthermore, if the layout
is bijective, the index mapping must involve a permutation of the shape
resulting from the tiling. This permutation can be inferred from a dimension
ordering of the strides $d$ (in ascending order). Consequently, the layout
inverse $H^{-1}$ must have shape $s'$ that is the inverse permutation of the
shape $s$. Furthermore, if $M_L^H$ is the layout mapping for $H$, the layout
mapping for $H^{-1}$ must be the inverse integer set relation $(M_L^H)^{-1}$.
Given the shape and overall layout mapping of the inverse layout, the required
layout inverse can be computed as described in Section~\ref{cute-layout-from-layout-mappings-and-shape}. 
This approach to obtain the layout inverse is summarized in Algorithm
\ref{alg:compute-inverse}.

\paragraph{Proof for correctness of Algorithm~\ref{alg:compute-inverse}:} For
bijective layout $H = s : d$, by definition the mapping is one-to-one and the
range has no gaps (size = co-size = $P$). The gap-free constraint is highly
restrictive as the index mapping $\sum_i d_i \times c_i$ must map the coordinate
space to the contiguous range $[0,P-1]$. This forces the strides $d$ to have a
permutation structure i.e., they must reorder the tiled coordinate dimensions
without creating holes. This is akin to tiling a 1-D loop-nest (coordinate
mapping), then some loop permutation followed by flattening it into 1-D
loop-nest again (which effectively permutes the colexicographic strides) to
create a different iteration order. The algorithm exploits this by: (1)
extracting the dimension permutation via stride ordering, (2) applying the
inverse permutation to reorder shape components, and (3) verifying that
$(M_L^H)^{-1} \circ (M_C^{H^{-1}})^{-1}$ yields a strictly affine relation
(enabling stride extraction via Algorithm 2).  The gap-free property of
bijective layouts ensures this verification succeeds.

Consider the bijective layout $H = (4, 2, 2) : (2, 1, 8)$. From the strides, we
get the dimension ordering $(1, 0, 2)$ which is the permutation involved in the
layout mapping. Its inverse permutation is $(1, 0, 2)$. Applying the inverse
permutation to  the shape of $H$ gives the shape of the inverse layout as $(2,
4, 2)$. From the inverse of the layout mapping $(M_L^H)^{-1}$ and the
coordinate mapping $M_C^H$, we can compute the index mapping of the inverse
layout, which is as follows:
\begin{align}
M_I^{H^{-1}} = \{ [c_0, c_1, c_2] \rightarrow [(4c_0 + c_1 + 8c_2)] : 0 \leq c_0 \leq 1 \land 0 \leq c_1 \leq 3 \land 0 \leq c_2 \leq 1 \}. \notag
\end{align}

The index mapping is strictly affine and the coefficients $(4, 1, 8)$
give us the strides of the inverse layout $H^{-1} = (2, 4, 2) : (4, 1, 8)$.


\footnotesize
\begin{mdframed}[style=algorithmstyle]
\begin{algorithm}[H]
\caption{Compute Inverse of Bijective Layout}
\label{alg:compute-inverse}
\begin{algorithmic}[1]

\Require $h$: bijective layout
\Ensure $h_{inv}$: inverse layout

\Function{compute\_inverse}{$h$}
    \State $s \gets \text{flatten\_tuple}(h.shape)$ \Comment{Flattened shape}
    \State $d \gets \text{flatten\_tuple}(h.strides)$ \Comment{Flattened strides}
    \State $p \gets \text{get\_dim\_order}(d)$ \Comment{Get dimension ordering based on ascending order of strides}
    \State $p_{inv} \gets \text{inverse\_permutation}(p)$
    \State $shape_{inv} \gets \langle s[i] \mid i \in p_{inv} \rangle$ \Comment{Permute shape}
    \State $l_{map} \gets l.\text{get\_layout\_mapping}().\text{reverse}()$  \Comment{reverse layout mapping}
    \State $c \gets \text{get\_coord\_mapping}(shape_{inv})$ \Comment{get coordinate mapping}
    \State $i \gets c.\text{reverse}().\text{apply\_range}(l_{map})$  \Comment{get index mapping}
    \State \Return $\text{get\_layout\_strictly\_affine}(i, shape_{inv})$ \Comment{Algorithm~\ref{alg:get-layout-strictly-affine}}
\EndFunction

\end{algorithmic}
\end{algorithm}
\end{mdframed}
\normalsize

Consider the layout $H=(2, 2):(1, 8)$ whose layout mapping $M_L^H$ is given by
\begin{equation}
    \label{eqn:cute-inverse-with-gaps}
M_L^H = \{ [c] \rightarrow [(-3 + 4c + 3((1 + c) \bmod 2))] : 0 \leq c \leq 3 \}.
\end{equation}
The layout mapping relation is bijective. However, $(M_L^H)^{-1}$ cannot be
represented as a CuTe layout due to gaps in its domain. So, 
a right inverse $H^r$ can be computed by finding the beginning point
\textit{begin} of the lexicographically smallest gap in the range of $M_L^H$.
By intersecting the integer set corresponding to the interval $[0, begin)$ with
the range of $M_L^H$, we obtain a relation without gaps that can be used to
determine the right inverse. Specifically, we need to find a bijective CuTe
layout $K = s : d$ whose layout mapping equals this relation. For this
example, the relation is as follows:
\begin{align}
M^\prime_L = \{ [c] \rightarrow [(c)] : 0 \leq c \leq 1 \}. \notag
\end{align}

To determine K, we intersect the domain of the coordinate mapping $M_C^{H}$ with
the domain of $M_L^K$ which gives us the relation
\begin{align}   
M^\prime_C =  \{ [c] \rightarrow [c0, c1] : c0 = c \land c1 = 0 \land 0 \leq c \leq 1 \} \notag.
\end{align}
The range of this relation can be used to extract a shape $s$. 
The composition $M'_L \circ (M'_C)^{-1}$ gives us a
strictly affine index mapping that can be used to determine the strides $d$
using Algorithm \ref{alg:get-layout-strictly-affine}. Finally, if the layout $K
= s : d$ is bijective, its inverse gives the required right inverse $H^r$. For
this example, we get $K = 2 : 1$, which is its own inverse. If the beginning
point of the gap is 1 or if the relation $M'_L$ is not bijective, then the right
inverse is given by the trivial layout $(1, 0)$.

Similarly, to compute the left inverse $H^l$ of $H$, we compute the layout
complement $H'$ of $H$ with respect to the target size $||H||$. This effectively
fills up the gaps in the range of $M_L^H$. The left inverse $H^l$ is obtained by
computing the right inverse of the concatenated layout $HH'$.

\section{CuTe Swizzles as Integer Set Relations}
\label{sec:cute-swizzles}
\begin{figure}[t]
    \centering
    \includegraphics[width=0.6\textwidth]{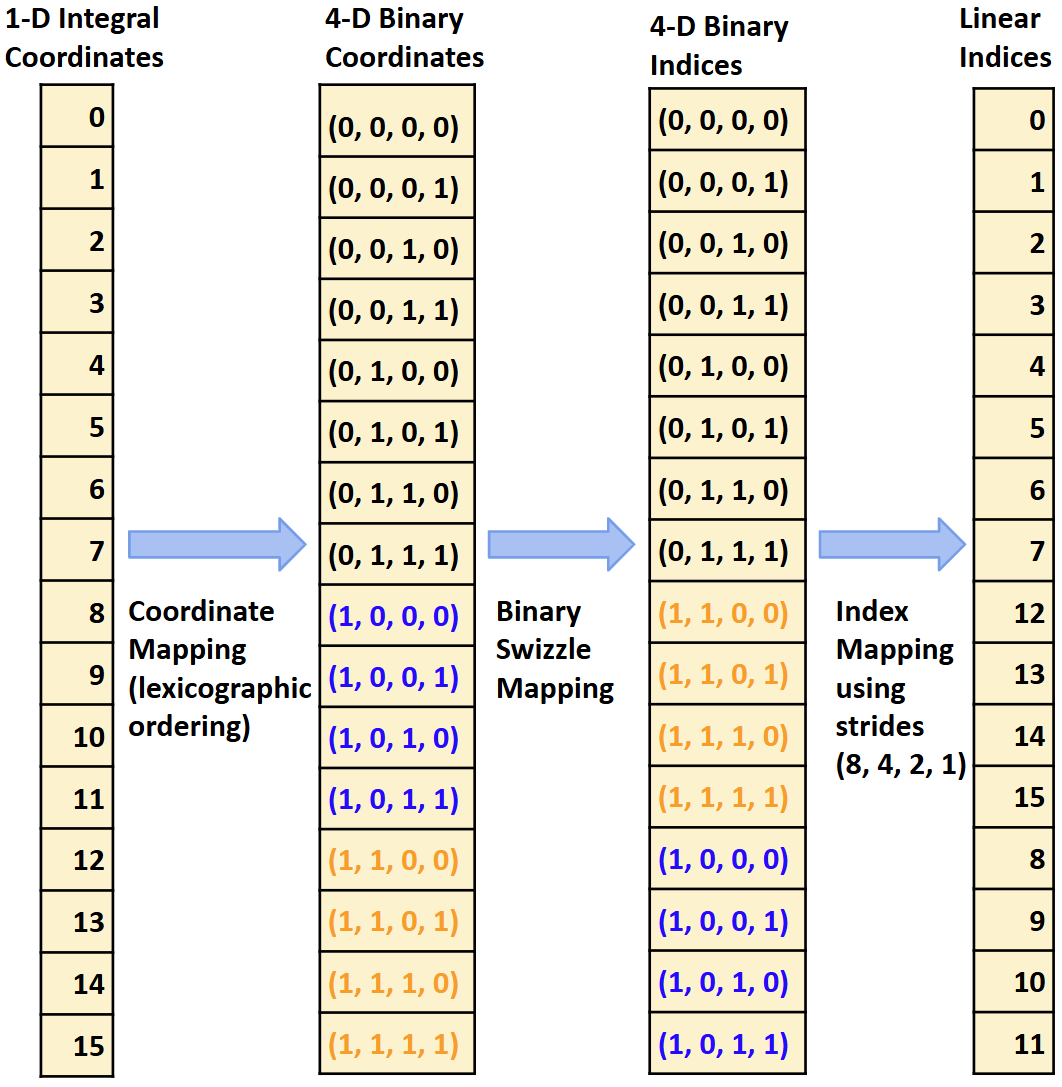}
    \caption{Layout mapping for CuTe swizzle $H_{1, 2, 1}$}
    \label{fig:swizzle-mapping}
\end{figure}
This section describes how CuTe swizzles can be expressed as integer set relations using ISL.

\subsection{Coordinate and Index mapping}
As defined in Section~\ref{sec:cute-swizzle}, a CuTe swizzle $H_{b, m, s}$
defines a mapping from an integral coordinate space of size $2^{b + m + |s|}$ to
a linear index space of the same size.  Since CuTe swizzles operate on binary
representations of coordinates, given a swizzle $H_{b, m, s}$ with $n = b + m +
|s|$, each coordinate $c$ in the integral coordinate space of size $2^n$ can be
mapped to an $n$-D binary coordinate space where each dimension has
size 2.  Similarly, each index in the linear index space can be mapped to a
distinct integer point in an $n$-D binary index space where each dimension is of
size 2. For example, consider the CuTe swizzle $H_{1, 2, 1}$ which defines a
mapping from an coordinate space of size $2^{1 + 2 + 1} = 16$ to a linear index
space of the same size. Equivalently, it can be seen as defining a mapping from
a 4-D binary coordinate space to 4-D linear index space. The exact mapping is
as shown in Figure \ref{fig:swizzle-mapping}.


The coordinate mapping from the integral coordinate space to the binary
coordinate space is based on a lexicographic ordering of the integer points. In
general, the coordinate mapping is given by the integer set relation:
\begin{align}
    M_R = &\{[c] \rightarrow [\lfloor c/2^{n-1} \rfloor, \lfloor c/2^{n-2} \rfloor \bmod 2, \lfloor c/2^{n-3} \rfloor \bmod 2, \ldots, c \bmod 2] : 0 \leq c < 2^n\}.
\end{align}

Comparing this to the coordinate mapping based on a colexicographic ordering
of integer points described in Section~\ref{coordinate-mapping}, it can be seen
that this is akin to a tiling transformation using tile sizes that are all equal
to 2. Furthermore, note that the mapping from an $n$-D binary index space to the
linear index space is also based on lexicographic ordering and so, is given by
the inverse $(M_R)^{-1}$ of the above integer set relation.
\begin{mdframed}[style=algorithmstyle]
\footnotesize
\begin{algorithm}[H]
\caption{Binary Swizzle Mapping for given CuTe Swizzle}
\label{alg:binary-swizzle-mapping}
\begin{algorithmic}[1]
\Require $b, m, s$ \Comment{Swizzle parameters}
\Ensure $M$ \Comment{integer set relation from binary coordinate space to binary index space}

\State $n \leftarrow b + m + |s|$ \Comment{Total dimensions}
\State $y \leftarrow ((2^b - 1) \ll (m + \max(s, 0)))$ \Comment{Mask value}
\State $y_{tuple} \leftarrow$ binary representation of $y$ padded to $n$ bits

\If{$s \geq 0$} \Comment{Positive shift}
    \For{$j = 0$ to $n - 1$}
        \If{$j - s \geq 0$}
            \State $expr_j \leftarrow i_j = (c_{j-s} \cdot y_{tuple}[j-s] + c_j) \bmod 2$
        \Else
            \State $expr_j \leftarrow i_j = c_j$
        \EndIf
    \EndFor
\Else \Comment{Negative shift}
    \For{$j = 0$ to $n - 1$}
        \If{$j - s < n$}
            \State $expr_j \leftarrow i_j = (c_{j-s} \cdot y_{tuple}[j-s] + c_j) \bmod 2$
        \Else
            \State$expr_j \leftarrow i_j = c_j$
        \EndIf
    \EndFor
\EndIf

\State $constraints \leftarrow expr_0 \land expr_1 \land \ldots \land expr_{n-1}$
\State $bounds \leftarrow \bigwedge_{i=0}^{n-1} (c_i \geq 0 \land c_i \leq 1)$
\State $M \leftarrow \{(c_0, \ldots, c_{n-1}) \rightarrow (i_0, \ldots, i_{n-1}) : constraints \land bounds\}$

\State \Return $M$
\end{algorithmic}
\end{algorithm}
\normalsize
\end{mdframed}

\subsection{Binary Swizzle Mapping}
\label{sec:binary-swizzle-mapping}
The CuTe swizzle is defined as a mapping from each point $c$ in the integral
coordinate space to a linear index $i = c \oplus ((c \land y) \gg s)$ where $y =
((2^b - 1) \ll (m + \max(s, 0)))$. Assuming positive shift width, we can make
the following observations:

\begin{itemize}
    \item The mask value $y$ is used to mask out the $m + s$ least significant
    bits of $c$. The result of this masking is then shifted by $s$ bits before it
    is XOR'ed with $c$.  So, the $m$ least significant bits in both $c$ and $i$
    must be identical. 

    \item Due to the shift operation, the next $b$ bits in $i$ would be computed
    as an XOR operation involving corresponding bits in $c$ and the bit value in
    $c$ that is $s$ bits away. 

    \item All other bits in $i$ are identical to their corresponding bits in $c$.
 \end{itemize}

Essentially, $b$ bits in $i$ are XOR'ed with bits $s$ positions away. The $m$
least significant bits remain unchanged due to the masking operation.  These
observations can be used to arrive at an integer set relation that captures the
swizzle mapping from the binary coordinate space to the binary index space.
Even in Figure~\ref{fig:swizzle-mapping} for a swizzle $H_{1, 2, 1}$, it can be
seen that the values in the last 2 dimensions of the binary coordinates and
indices are identical. So, the swizzle mapping $M_{S}^H$ is given by:
\begin{align}
    M_{S}^H =& \{ [c_0, c_1, c_2, c_3] \rightarrow [c_0, (c_0 + c_1) \bmod 2, c_2, c_3] : 0 \leq c_k \leq 1 \text{ for } k \in [0, 3] \}.
\end{align}
Due to the shift width being equal to 1, the second dimension is computed as an
XOR of the values $c_1$ and $c_0$ (the XOR operation is expressed using mod 2). In
general, assuming positive shift width, this can be generalized to define the
swizzle mapping $M_{S}^H$ for a CuTe swizzle $H_{b, m, s}$ as follows, where
$n = b + m + |s|$:
\begin{align}
    M_{S}^H =& \{ [c_0, \ldots, c_{n-1}] \rightarrow [c_0, \ldots, c_{s-1}, (c_0 + c_s) \bmod 2, \ldots, (c_{n-1-m-s} + c_{n-1-m}) \bmod 2, \notag \\
             & c_{n-1-m+1}, \ldots, c_{n-1}] : 0 \leq c_k \leq 1 \text{ for } k \in [0, n-1] \}.
\end{align}

It can be seen that the binary swizzle mapping is an involution
as $M_{S}^H \circ M_{S}^H = I$ where $I$ is the identity mapping.
Algorithm~\ref{alg:binary-swizzle-mapping} shows an algorithm to obtain such a
swizzle mapping for the CuTe swizzle $H_{b, m, s}$. Note that the shift value
can be positive or negative indicating a right or left shift and the algorithm
accounts for this.
\paragraph{Proof for correctness of Algorithm~\ref{alg:binary-swizzle-mapping}:} The algorithm essentially 
implements the swizzle formula $i = c \oplus ((c \wedge y) \gg s)$ in binary
space. The mask $y$ isolates the $b$ bits to be XOR'ed, and the shift operation
correctly positions them. For positive shift, bits $j-s$ through $j-s+b-1$ are
XOR'ed with bits $j$ through $j+b-1$. For negative shift, the roles reverse. The
modulo-2 arithmetic correctly implements XOR, and the bounds ensure all
variables remain binary.

\subsection{Layout Mapping}

Overall, the layout mapping defined by the CuTe swizzle from the 1-D integral
coordinate space to the 1-D index space can be obtained by the composition of
the three mappings discussed above in that order -- coordinate mapping, swizzle
mapping and index mapping. 
\begin{align}
    M_L^H = (M_R)^{-1} \circ M_S \circ M_R
\end{align}

The symmetry of the composition above shows that the layout mapping is an
involution i.e., it is its own inverse. The layout mapping for the CuTe swizzle
$H_{1, 2, 1}$ is as follows:
\begin{align}
    M_L^H = & \{ [c] \rightarrow [(c - (c) \bmod 8 + (c + 4\lfloor(c/8)\rfloor) \bmod 8)] : 0 \leq c \leq 15 \}.
\end{align}

\section{Linear Layouts as Integer Set Relations}
\label{sec:ll}
\label{layout-mappings-as-integer-set-set-relations}

This section describes how various mappings defined by a Triton linear layout
can be systematically derived and represented using ISL integer set relations.

\subsection{Coordinate and Index Mapping}
\label{coordinate-and-index-mapping}

As introduced in Section~\ref{sec:linear-layout}, a linear layout $H$ defines a
mapping from an $m$-D coordinate space to an $n$-D index space by specifying the
mappings for the basis vectors of the coordinate space. We will refer to these
spaces as the natural coordinate and natural index spaces respectively. The
layout is characterized by three components: the natural coordinate space shape,
the natural index space shape, and the mapping values for basis vectors. For
example, consider the linear layout $H = (\text{crd}=(4, 4), \text{idx}=(4, 4),
\text{vals}=[(1, 1), (2, 2), (0, 1), (0, 2)])$.  This layout specifies a 2-D
coordinate space and a 2-D index space both of shape $(4, 4)$.  The four basis
vectors of the coordinate space in colexicographic order are $(1, 0)$, $(2, 0)$,
$(0, 1)$, and $(0, 2)$ and they are mapped to indices $(1, 1)$, $(2, 2)$, $(0,
1)$, and $(0, 2)$. 

As described in Section~\ref{sec:linear-layout}, $M$-D and $N$-D
binary vector spaces can be used to represent the coordinate and index spaces
respectively. Note that $M$ is also the number of basis vectors in the
coordinate space. So, in the example above, $M = 4$ and $N = 4$
and the basis vectors in the binary coordinate space are $(1, 0, 0, 0)$, $(0,
1, 0, 0)$, $(0, 0, 1, 0)$, and $(0, 0, 0, 1)$. These are mapped to the points
$(1, 0, 1, 0)$, $(0, 1, 0, 1)$, $(0, 0, 1, 0)$, and $(0, 0, 0, 1)$ in the binary
index space.  Any coordinate can be expressed as a linear combination of these
basis vectors.  The linearity rule shown in
Equation~\ref{eq:linear-layout-transformation}  is used to
compute the corresponding index in the binary index space.

\begin{figure}[t]
    \centering
    \includegraphics[width=0.9\textwidth]{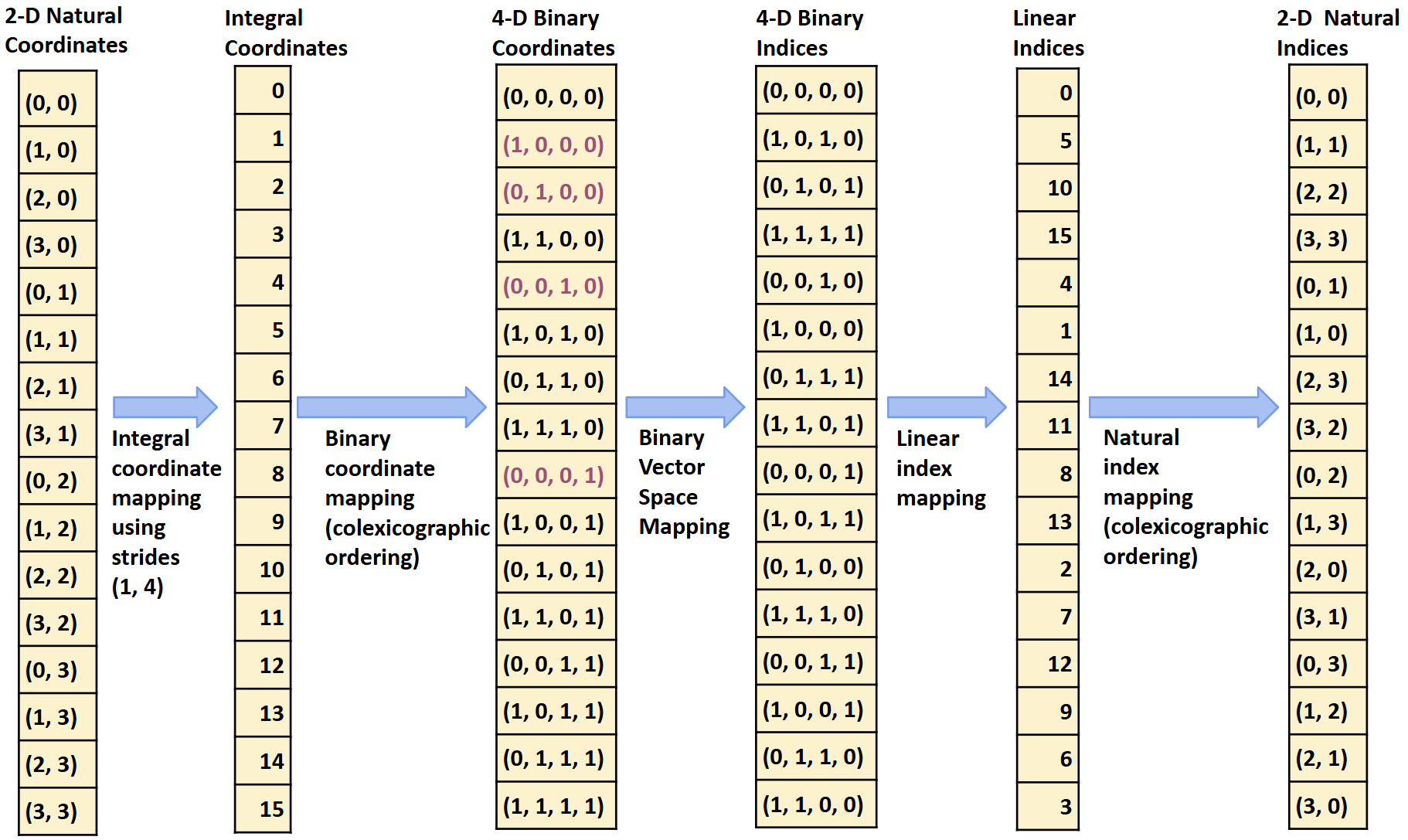}
    \caption{Linear layout mapping for $H=(crd=(4,4), idx=(4,4), vals=[(1,1), (2,2), (0,1), (0,2)])$}
    \label{fig:ll-mapping}
\end{figure}

As shown in Figure~\ref{fig:ll-mapping}, the complete linear layout mapping from the
$m$-D natural coordinate space to $n$-D natural index space can be derived as the
composition of the following mappings.
\begin{enumerate}
    \item An integral coordinate mapping $M_{ic}^H$ that linearizes the $m$-D natural coordinates
          to 1-D integral coordinates based on the colexicographic ordering of the coordinates. 
    \item A binary coordinate mapping $M_{bc}^H$ that maps the 1-D integral coordinates to
          $M$-D binary coordinates based on colexicographic ordering.
    \item Binary vector space mapping $M_{bv}^H$ that maps the $M$-D binary coordinates to
          $N$-D binary coordinates.
    \item A linear index mapping $M_{li}^H$ that linearizes the $N$-D binary indices to
          1-D linear indices based on the colexicographic strides $1, 2, 2^2, \ldots, 2^{N-1}$.
    \item A natural index mapping $M_{ni}^H$ that maps the 1-D linear index space to
          $n$-D natural index space using colexicographic ordering.
\end{enumerate}

For a linear layout $H$, suppose  the $m$-D natural coordinate space is of shape
$(s_0, s_1, \ldots, s_{m-1})$, and the $n$-D natural index space is of shape
$(s'_0, s'_1, \ldots, s'_{n-1})$.  Consequently, the sizes of the $M$-D binary
coordinate space and the $N$-D binary index space are $M = \log s_0
+ \log s_1 + \ldots + \log s_{m-1}$ and $N = \log s'_0 + \log s'_1 + \ldots + \log
s'_{n-1}$ respectively, with each dimension of the binary spaces being of
size 2.

The integer set relations for the various mappings are as shown below.

\begin{equation}
M_{ic}^H = \{[c_0, \ldots, c_{m-1}] \rightarrow [c] : \forall i \in [0, m), 0 \leq c_i < s_i \land c = \sum_{i=0}^{m-1} c_i \prod_{j=0}^{i-1} s_j\}
\end{equation}
\begin{equation}
M_{bc}^H = \{[c] \rightarrow [c_0, \ldots, c_{M-1}] : \forall i \in [0, M), 0 \leq c_i < 2 \land c = \sum_{i=0}^{M-1} c_i 2^i\}
\end{equation}
\begin{equation}
M_{li}^H = \{[i_0, \ldots, i_{N-1}] \rightarrow [i] : \forall j \in [0, N), 0 \leq i_j < 2 \land i = \sum_{j=0}^{N-1} i_j 2^j\}
\end{equation} 
\begin{equation}
M_{ni}^H = \{[i] \rightarrow [i_0, \ldots, i_{n-1}] : \forall j \in [0, n), 0 \leq i_j < s'_j \land i = \sum_{j=0}^{n-1} i_j \prod_{k=0}^{j-1} s'_k\}
\end{equation}
\subsection{Mapping Between Binary Vector Spaces}
\label{mapping-between-binary-vector-spaces}
In order to completely define the linear layout $H$, $M$ natural index values
are specified and these correspond to the $M$ basis vectors of the natural coordinate
space. Now, consider the inverse $(M_{ni}^H \circ M_{li}^H)^{-1}$. This maps the
$n$-D natural index space to the N-D binary index space and can be used to
obtain the corresponding $M$ binary index values.  Suppose these are $d_0, d_1,
\ldots, d_{M-1}$. Each $d_j = (d_j^0, d_j^1, \ldots, d_j^{N-1})$ represents the
index value for the $j$-th basis vector.

The index value $i = (i_0, i_1, \ldots, i_{N-1})$ for an arbitrary point $c =
(c_0, c_1, \ldots, c_{M-1})$ in the binary coordinate space is computed as follows:
\begin{align}
i_j = d_0^j \cdot c_0 \oplus d_1^j \cdot c_1 \oplus \ldots \oplus d_{M-1}^j \cdot c_{M-1}.
\end{align}
Since the linear mapping is between
binary vector spaces, multiplication is replaced by logical AND and addition by
logical XOR. Logical XOR of $a$ and $b$ is equivalent to $(a + b) \bmod 2$ and
logical AND can just be treated as multiplication. So, we can rewrite the above
as follows:
\begin{align}
i_j =  (d_0^j c_0 + d_1^j c_1 + \ldots + d_{M-1}^j c_{M-1}) \bmod 2.
\end{align}

The binary vector space mapping $M_{bv}^H$ is given by the integer set relation:
\begin{align}
M_{bv}^H = &\{ [c_0, \ldots, c_{M-1}] \rightarrow [i_0, \ldots, i_{N-1}] : 0 \leq c_k < 2 \text{ for } k \in [0, M-1] \notag \\
&\land i_j = \sum_{k=0}^{M-1} d_k^j c_k \bmod 2 \text{ for } j \in [0, N-1]\}.
\end{align}

\subsection{Layout Mapping}
\label{ssec:layout-mapping}

The complete linear layout mapping from $m$-D coordinate space to
$n$-D index space is expressed as the composition of all derived
integer set relations:
\begin{equation}
    M_L^H = M_{ni}^H \circ M_{li}^H \circ M_{bv}^H \circ M_{bc}^H \circ M_{ic}^H.
\end{equation}

For the example linear layout $H=(crd=(4,4), idx=(4,4), vals=[(1,1),
(2,2), (0,1), (0,2)])$, the binary vector space mapping $M_{bv}^H$ is given by
the following integer set relation.  
\begin{equation}
\{[c_0, c_1, c_2, c_3] \rightarrow [ c_0, c_1, (c_0 + c_2) \bmod 2, (c_1 + c_3) \bmod 2] : 0 \leq c_0, c_1, c_2, c_3 \leq 1\}.
\end{equation}

Note that from the form of this integer set relation, it is clear that it
represents a swizzled layout that is equivalent to the CuTe swizzle $H_{2,0,-2}$
(recall that CuTe swizzle is based on lexicographic ordering while linear layout
is based on colexicographic ordering). This showcases the expressive power of
integer set relations in representing diverse layout abstractions while
providing a unified framework for comparing and reasoning across different
layout abstractions.

Finally, the overall linear layout mapping from the 2-D coordinate space to 2-D
index space, computed by ISL using the decomposition described above is as
follows:
\begin{align}
\{ [c_0, c_1] &\rightarrow [(c_0), (3 + (c_0) \bmod 2 - (1 + c_0 + c_1) \bmod 2 - (3 + c_0 + 3c_1 - (1 + c_1) \bmod 2) \bmod 4)]  \notag \\
            &: 0 \leq c_0 \leq 3 \land 0 \leq c_1 \leq 3\}.
\end{align}

\section{Implementation and Examples}
\label{sec:implementation}
\begin{center}
\footnotesize
\begin{table*}[t]
\begin{tabular}{|p{0.25\textwidth}|p{0.7\textwidth}|}
\hline
\textbf{Layout/Mapping Type} & \textbf{Representation} \\
\hline

\textbf{swizzled} & $LinearLayout( crd=(4, 4), idx=(4, 4), vals=[(1, 1), (2, 2), (0, 1), (0, 2)])$  \\
Binary Vector Space Mapping & $\{ [c_0, c_1, c_2, c_3] \rightarrow [(c_0), (c_1), ((c_0 + c_2) \mod 2), ((c_1 + c_3) \mod 2)] : 0 \leq c_0,c_1,c_2,c_3 \leq 1 \}$  \\
Layout Mapping & $\{ [c_0, c_1] \rightarrow [(c_0), (1 - (c_0) \mod 2 - (1 + c_0 + c_1) \mod 2 + (1 + c_0 + 3c_1 - (1 + c_1) \mod 2) \mod 4)] : 0 \leq c_0,c_1 \leq 3 \}$ \\
\hline

\textbf{1d\_identity} & $LinearLayout( crd=8, idx=8, vals=[1, 2, 4])$  \\
Binary Vector Space Mapping & $\{ [c_0, c_1, c_2] \rightarrow [(c_0), (c_1), (c_2)] : 0 \leq c_0,c_1,c_2 \leq 1 \}$  \\
Layout Mapping & $\{ [c_0] \rightarrow [(c_0)] : 0 \leq c_0 \leq 7 \}$ \\
\hline

\textbf{zeros} & $LinearLayout( crd=8, idx=8, vals=[0, 0, 0])$  \\
Binary Vector Space Mapping & $\{ [c_0, c_1, c_2] \rightarrow [(0)] : 0 \leq c_0,c_1,c_2 \leq 1 \}$  \\
Layout Mapping & $\{ [c_0] \rightarrow [(0)] : 0 \leq c_0 \leq 7 \}$ \\
\hline

\textbf{2d\_identity} & $LinearLayout( crd=(4, 4), idx=(4, 4), vals=[(1, 0), (2, 0), (0, 1), (0, 2)])$  \\
Binary Vector Space Mapping & $\{ [c_0, c_1, c_2, c_3] \rightarrow [(c_0), (c_1), (c_2), (c_3)] : 0 \leq c_0,c_1,c_2,c_3 \leq 1 \}$  \\
Layout Mapping & $\{ [c_0, c_1] \rightarrow [(c_0), (c_1)] : 0 \leq c_0,c_1 \leq 3 \}$ \\
\hline

\textbf{2d\_transpose} & $LinearLayout( crd=(4, 4), idx=(4, 4), vals=[(0, 1), (0, 2), (1, 0), (2, 0)])$  \\
Binary Vector Space Mapping & $\{ [c_0, c_1, c_2, c_3] \rightarrow [(c_2), (c_3), (c_0), (c_1)] : 0 \leq c_0,c_1,c_2,c_3 \leq 1 \}$  \\
Layout Mapping & $\{ [c_0, c_1] \rightarrow [(c_0), (c_1)] : 0 \leq c_0,c_1 \leq 3 \}$ \\
\hline

\textbf{1d\_transpose} & $LinearLayout( crd=16, idx=16, vals=[4, 8, 1, 2])$  \\
Binary Vector Space Mapping & $\{ [c_0, c_1, c_2, c_3] \rightarrow [(c_2), (c_3), (c_0), (c_1)] : 0 \leq c_0,c_1,c_2,c_3 \leq 1 \}$  \\
Layout Mapping & $\{ [c_0] \rightarrow [(15 + 4c_0 + 15\lfloor((-1 - c_0)/4)\rfloor)] : 0 \leq c_0 \leq 15 \}$ \\
\hline

\textbf{2d\_broadcast} & $LinearLayout( crd=(4, 4), idx=4, vals=[1, 2, 0, 0])$  \\
Binary Vector Space Mapping & $\{ [c_0, c_1, c_2, c_3] \rightarrow [(c_0), (c_1)] : 0 \leq c_0,c_1,c_2,c_3 \leq 1 \}$  \\
Layout Mapping & $\{ [c_0, c_1] \rightarrow [(c_0)] : 0 \leq c_0,c_1 \leq 3 \}$ \\
\hline
\end{tabular}
\caption{Linear Layouts and their equivalent integer set relations}
\label{tab:linear-layouts-and-equivalent-integer-set-relations}
\end{table*}
\normalsize
\end{center}

We developed \texttt{isl-layout}, a tool that implements the techniques
described in previous sections to translate CuTe and linear layout
specifications into integer set relations using ISLpy~\cite{islpy}, a python
wrapper for ISL. The tool supports all the CuTe operations described in
Section~\ref{layout-operations} thereby providing a unified framework for
analyzing and manipulating layout abstractions. We chose ISL for its
comprehensive support for integer sets and relations, including those with
quasi-affine constraints, and proven scalability in production compiler systems
(LLVM Polly, GCC Graphite). We also preferred ISL for its mature Python support
through ISLpy. 

Table~\ref{tab:linear-layouts-and-equivalent-integer-set-relations} demonstrates
the translation of various linear layouts into equivalent integer set relations.
The examples are drawn from the Triton codebase and showcase the
ability of the framework to handle diverse layout patterns.  The table shows three key components
for each layout -- the linear layout definition, the binary vector space
mapping and the overall layout mapping.

Table~\ref{tab:layout_operations} presents CuTe layouts and operations with
their corresponding integer set relations. The table demonstrates the capability
to handle complex multi-dimensional layouts with non-trivial stride patterns,
various operations such as layout inversion, complement and composition, as well
as CuTe swizzled layouts.


\paragraph{Expressiveness Beyond Existing Layout Systems} Integer set relations
encompass a richer class of layout mappings beyond what CuTe and linear layouts
can express. Mappings involving arbitrary non-rectangular integer sets (e.g.,
triangular domains) or non-convex integer set domains with holes represent clear
examples beyond existing layout system capabilities. Even with rectangular
integer sets, certain complex shuffles exceed the representational capacity of
both systems. For instance, consider the shuffle $S = \{[c_0] \to [((-3c_0)
\bmod 16)] : 0 \leq c_0 \leq 15\}$, which represents a kind of rotational and
coprime fanning permutation discussed by Phothilimthana et
al.~\cite{phothilimthana2019swizzle}. This modulo mapping cannot be represented
by existing layout systems as it is neither strided nor based on binary swizzle
mappings. Overall, integer set relations expand the layout optimization search
space significantly, opening up opportunities for exploring more sophisticated
memory access patterns.

\paragraph{Complexity} Translation from a given layout system to
integer set relations is a compile-time processing step and not a runtime
operation. Although integer set operations like relation composition and
lexicographic minimum exhibit worst-case exponential complexity, ISL's design
optimizes polyhedral analysis for practical use. Its widespread deployment in
production compilers like LLVM Polly demonstrates negligible computational
overhead in real-world scenarios.  For example, most tensors in deep learning
networks like Torchvision and Hugging Face Transformers have rank $\leq 6$. Even
with a 4-level tiling commonly used in GPU kernel optimization, there would be
at most 24 dimensions or variables, which is well within the efficient operating
range of ISL.

\begin{center}
\footnotesize
\begin{table*}[t]
\centering
\begin{tabular}{|p{0.25\textwidth}|p{0.4\textwidth}|p{0.27\textwidth}|}
\hline
\textbf{CuTe layout/operation} & \textbf{Layout mapping} & \textbf{Resulting layout} \\
\hline
$(4, 2, 2):(2, 1, 8)$ & $\{ [c] \rightarrow [(7 + 2c + 6\lfloor(c/8)\rfloor + 7\lfloor(-1 - c)/4\rfloor)] : 0 \leq c \leq 15 \}$ & \\
\hline
inverse$((4, 2, 2):(2, 1, 8))$ & $\{ [c] \rightarrow [(-3c + 4\lfloor(c/8)\rfloor + 7\lfloor(1 + c)/2\rfloor)] : 0 \leq c \leq 15 \}$ & $(2, 4, 2):(4, 1, 8)$ \\
\hline
left\_inverse$((4, 2, 2):(4, 2, 32))$ & $\{ [c] \rightarrow [(2c - 7\lfloor(c/4)\rfloor + 28\lfloor(c/16)\rfloor - 56\lfloor(c/32)\rfloor + 14((c) \bmod 2))] : 0 \leq c \leq 63 \}$ & $(2, 2, 4, 2, 2):(16, 4, 1, 32, 8)$ \\
\hline
right\_inverse$(4, 8, 2):(8, 1, 33))$ & $\{ [c] \rightarrow [(31 + 4c + 31\lfloor((-1 - c)/8)\rfloor)] : 0 \leq c \leq 31 \}$ & $(8, 4):(4, 1)$ \\
\hline
complement$((2, 2):(1, 5), 20)$ & $\{ [c] \rightarrow [(2c + 5\lfloor(c/2)\rfloor)] : 0 \leq c \leq 5 \}$ & $(2, 3):(2, 9)$ \\
\hline
complement$((4, 2):(1, 16), 32)$ & $\{ [c] \rightarrow [(4c)] : 0 \leq c \leq 3 \}$ & $4 : 4$ \\
\hline
complement$((2, 2):(2, 10), 20)$ & $\{ [c] \rightarrow [(-1 + 2c + 10\lfloor(c/4)\rfloor + ((1 + c) \bmod 2))] : 0 \leq c \leq 7 \}$ & $((2, 2), 2):((1, 4), 18)$ \\
\hline
complement$((2, 2):(1, 4), 20)$ & $\{ [c] \rightarrow [(-2 + 4c + 2((1 + c) \bmod 2))] : 0 \leq c \leq 5 \}$ & $(2, 3):(2, 8)$ \\
\hline
$(2, 2):(1, 80) \circ (2, 2):(2, 1)$ & $\{ [c] \rightarrow [(-79c + 159\lfloor(1 + c)/2\rfloor)] : 0 \leq c \leq 3 \}$ & $(2, 2):(80, 1)$ \\
\hline
$(4, 6, 8, 10):(2, 3, 5, 7) \circ 6 : 12$ & $\{ [c] \rightarrow [(-4c + 13\lfloor(1 + c)/2\rfloor)] : 0 \leq c \leq 5 \}$ & $(2, 3):(9, 5)$ \\
\hline
$(4, 2, 8):(3, 12, 97) \circ 3 : 3$ & $\{ [c] \rightarrow [(9c)] : 0 \leq c \leq 2 \}$ & $ 3 : 9$ \\
\hline
\parbox[t]{4cm}{$((4, 2),(2, 4)):((2, 16), (1, 8))$ \\ $\circ ((4, 8), 2):((16, 1), 8)$} & 
\parbox[t]{5cm}{$\{ [c] \rightarrow [(30 + 8c + 8\lfloor(c/16)\rfloor - 31\lfloor(c/32)\rfloor + 30\lfloor(-1 - c)/4\rfloor)] : 0 \leq c \leq 63 \}$} & $(4, (4, 2)), 2):((8, (2, 16)), 1)$ \\
\hline
$(4, 2, 2):(2, 1, 8) \circ 16 : 1$ & $\{ [c] \rightarrow [(7 + 2c + 6\lfloor(c/8)\rfloor + 7\lfloor(-1 - c)/4\rfloor)] : 0 \leq c \leq 15 \}$ & $(4, 2, 2):(2, 1, 8)$ \\
\hline
swizzle$(1, 2, 1)$ & $\{ [c] \rightarrow [(c - (c) \bmod 8 + (c + 4\lfloor(c/8)\rfloor) \bmod 8)] : 0 \leq c \leq 15 \}$ & \\
\hline
swizzle$(1, 2, -1)$ & $\{ [c] \rightarrow [(-7 + 2((c) \bmod 8) + (7 + c - 2((c) \bmod 4)) \bmod 16)] : 0 \leq c \leq 15 \}$ & \\
\hline
\end{tabular}
\caption{CuTe layout/operations and their corresponding mappings}
\label{tab:layout_operations}
\end{table*}
\normalsize
\end{center}

\vskip -60pt
\section{Related Work}
\label{sec:related}
Our research intersects with established domains in compiler optimization, 
GPU computing, and deep learning systems. We describe our contributions within 
this broader context while highlighting the novel aspects of our approach.

The polyhedral model has emerged as a powerful framework for analyzing and
transforming loop nests across diverse application domains. Early applications
focused on scientific computing~\cite{uday08pldi} and high-performance
computing~\cite{polly}, with subsequent extensions to domain-specific
languages~\cite{baghdadi2019cgo, mullapudi2015asplos}. The Integer Set Library
(ISL)~\cite{isl} established the computational infrastructure for manipulating
integer sets and relations and is used in many polyhedral tools~\cite{pet,
clan_bastoul}. It also provides the foundation upon which our work builds.
Polyhedral compilation has also proven to be effective for GPU code
generation~\cite{verdoolaege2013taco}. These developments illustrate
the expressive power of integer relations used in polyhedral compilation,
motivating our application of its ideas to analyze layout abstractions.

Modern deep learning frameworks have increasingly adopted polyhedral techniques
to optimize neural network computations. Tensor
Comprehensions~\cite{vasilache2018tcCoRR} and Diesel~\cite{elango2018mapl}
demonstrated the effectiveness of polyhedral analysis for automatic kernel
generation, while MLIR~\cite{mlircgo2021} established a unified infrastructure for
compiler transformations.  GPU optimization for deep learning has evolved
through multiple generations of specialized libraries and frameworks.
cuDNN~\cite{chetlur2014arxiv} established the foundation with hand-tuned
primitives, while subsequent work explored automatic kernel
generation~\cite{bhaskaracharya2020arxiv} and domain-specific
languages~\cite{graphene_asplos}. 

Contemporary deep learning systems employ diverse layout abstraction mechanisms.
CuTe layouts~\cite{cute} in CUTLASS~\cite{cutlass} provide a comprehensive
system for expressing tensor layouts and transformations, while Triton linear
layouts~\cite{zhou2025linearlayoutsrobustcode, triton_linear_layout} offer an
alternative approach based on binary vector space mappings.
Tilus~\cite{tilus2025arxiv} provides a virtual machine abstraction for arbitrary
low-precision computation and employs a layout system that mostly supports
strided layouts, thereby being a strict subset of the CuTe layout system. These
systems demonstrate the practical importance of layout abstractions. Recent work
includes a categorical foundation for CuTe layouts~\cite{colfax2025categorical},
which establishes its connections to category theory and operads. This work
introduces a graphical calculus for computing layout operations. While the
categorical approach provides elegant theoretical foundations for understanding
CuTe layouts and their algebra, it operates at a different level of abstraction
from polyhedral methods and does not directly support a unified mathematical
framework for various existing layout abstractions such as swizzled and linear
layouts.

While existing work has established the importance of both polyhedral
compilation and layout abstractions in deep learning systems, no prior research
has systematically correlated these domains. To the best of our knowledge, our
work establishes the first unified mathematical framework that enables formal
analysis of layout mappings across different layout systems.  By modeling layout
abstractions as integer set relations, our approach enables their integration
into the FPL~\cite{pitchanathanOopsla2021, grosserOopsla2020}. This integration
would further facilitate the incorporation of layout abstractions into core
MLIR~\cite{mlircgo2021} dialects, making them accessible to most deep learning
compiler frameworks.

\section{Conclusions}
\label{sec:conclusions}



This paper presents a unified mathematical framework for modeling layout
abstractions through integer set relations, bridging the gap between CuTe
layouts (including sophisticated swizzle operations) and Triton linear layouts.
To the best of our knowledge, this work establishes the first mathematical
framework that enables formal reasoning about layout abstractions across
different systems, thereby providing a common language for analyzing layouts and
their transformations. 

Our implementation demonstrates that the framework successfully handles the full
spectrum of layout complexity, from simple 1D mappings to complex 4D tensor
layouts with advanced swizzle patterns based on bit-level manipulations. The
mathematical representation of layouts as integer set relations also opens up
possibilities for integration with existing polyhedral compilation
frameworks. While this work focuses on establishing mathematical foundations
rather than runtime performance optimization, it represents a crucial step
towards unifying layout abstraction systems in deep learning compilers, thereby providing
the theoretical foundation necessary for advanced optimization techniques that
will be essential for the next generation of deep learning compilers.

\bibliographystyle{ACM-Reference-Format}
\bibliography{bibfile}


\end{document}